\documentclass[sigconf,nonacm,pdfa]{acmart}

\usepackage[utf8]{inputenc}
\usepackage{amsmath,amsfonts}
\usepackage{amsthm}
\usepackage{textcomp}
\usepackage{color}
\usepackage{xcolor}
\usepackage{colortbl}
\usepackage{multirow}
\usepackage{array}
\usepackage{enumitem}
\usepackage{epstopdf}
\usepackage{booktabs}
\usepackage{threeparttable}
\usepackage{bm}
\usepackage{graphicx}
\usepackage{subfigure}
\usepackage{caption}
\usepackage{balance}

\usepackage{amsmath}    

\usepackage{graphicx}
\usepackage{stfloats}

\newtheoremstyle{mydefn}
{}{}
{\it}       
{0pt}       
{\bfseries} 
{:~}        
{0.25em}    
{}          
\theoremstyle{mydefn}

\newtheoremstyle{myexample}
{}{}
{}          
{0pt}       
{\bfseries} 
{:~}        
{0.25em}    
{}          
\theoremstyle{myexample}

\renewcommand{\paragraph}[1]{\smallskip\noindent\textbf{#1.}}
\renewcommand{\subparagraph}[1]{\smallskip\noindent\textbf{\underline{#1.}}}
\renewcommand{\sec}[1]{\textbf{\S#1}}
\newcommand{\keypoint}[1]{\textbf{\underline{#1}}}

\newcommand{\floor}[1]{\lfloor #1 \rfloor}

\makeatletter
\newif\if@restonecol
\makeatother

\usepackage[linesnumbered,ruled,vlined]{algorithm2e}
\usepackage{algpseudocode}

\SetKwComment{Comment}{$\triangleright$\ }{}
\SetKwRepeat{Do}{do}{while}

\newcommand\vldbdoi{XX.XX/XXX.XX}
\newcommand\vldbpages{XXX-XXX}

\newcommand\vldbyear{2020}
\newcommand\vldbauthors{\authors}
\newcommand\vldbtitle{\shorttitle} 

\newcommand\vldbpagestyle{plain} 

\begin{document}

\title{CTBench: Cryptocurrency Time Series Generation Benchmark}

\author{Yihao Ang}
\orcid{0009-0009-1564-4937}
\affiliation{
 \institution{National University of Singapore}
 \country{}
}
\email{yihao\_ang@comp.nus.edu.sg}

\author{Qiang Wang}
\orcid{0009-0007-8429-2520}
\affiliation{
 \institution{National University of Singapore}
 \country{}
}
\email{qwang@u.nus.edu}

\author{Qiang Huang}
\orcid{0000-0003-1120-4685}
\affiliation{
 \institution{Harbin Institute of Technology (Shenzhen)}
 \country{}
}
\email{huangqiang@hit.edu.cn}

\author{Yifan Bao}
\orcid{0009-0000-9672-0747}
\affiliation{
 \institution{National University of Singapore}
 \country{}
}
\email{yifan\_bao@comp.nus.edu.sg}

\author{Xinyu Xi}
\orcid{0009-0008-7454-8024}
\affiliation{
 \institution{National University of Singapore}
 \country{}
}
\email{xinyu\_xi@u.nus.edu}

\author{Anthony K. H. Tung}
\orcid{0000-0001-7300-6196}
\affiliation{
 \institution{National University of Singapore}
 \country{}
}
\email{atung@comp.nus.edu.sg}

\author{Chen Jin}
\orcid{0000-0001-9940-0757}
\affiliation{
 \institution{National University of Singapore}
 \country{}
}
\email{disjinc@nus.edu.sg}

\author{Zhiyong Huang}
\orcid{0000-0002-1931-7775}
\affiliation{
 \institution{National University of Singapore}
 \country{}
}
\email{huangzy@comp.nus.edu.sg}

\begin{abstract}
Synthetic time series are essential tools for data augmentation, stress testing, and algorithmic prototyping in quantitative finance. 
However, in cryptocurrency markets, characterized by 24/7 trading, extreme volatility, and rapid regime shifts, existing Time Series Generation (TSG) methods and benchmarks often fall short, jeopardizing practical utility.
Most prior work (1) targets non-financial or traditional financial domains, (2) focuses narrowly on classification and forecasting while neglecting crypto-specific complexities, and (3) lacks critical financial evaluations, particularly for trading applications.
To address these gaps, we introduce \textsf{CTBench}, the first comprehensive TSG benchmark tailored for the cryptocurrency domain. 
\textsf{CTBench} curates an open-source dataset from 452 tokens and evaluates TSG models across 13 metrics spanning 5 key dimensions: forecasting accuracy, rank fidelity, trading performance, risk assessment, and computational efficiency.
A key innovation is a dual-task evaluation framework: (1) the \emph{Predictive Utility} task measures how well synthetic data preserves temporal and cross-sectional patterns for forecasting, while (2) the \emph{Statistical Arbitrage} task assesses whether reconstructed series support mean-reverting signals for trading. 
We benchmark eight representative models from five methodological families over four distinct market regimes, uncovering trade-offs between statistical fidelity and real-world profitability.
Notably, \textsf{CTBench} offers model ranking analysis and actionable guidance for selecting and deploying TSG models in crypto analytics and strategy development.
\end{abstract}

\maketitle

\pagestyle{\vldbpagestyle}
\begingroup\small\noindent\raggedright\textbf{ACM Reference Format:}\\
\vldbauthors. \vldbtitle. ACM Conference, \vldbpages, \vldbyear.\\
\href{https://doi.org/\vldbdoi}{doi:\vldbdoi}
\endgroup
\begingroup
\renewcommand\thefootnote{}\footnote{\noindent
Permission to make digital or hard copies of all or part of this work for personal or classroom use is granted without fee provided that copies are not made or distributed for profit or commercial advantage and that copies bear this notice and the full citation on the first page. Copyrights for components of this work owned by others than ACM must be honored. Abstracting with credit is permitted. To copy otherwise, or republish, to post on servers or to redistribute to lists, requires prior specific permission and/or a fee. Request permissions from permissions@acm.org. \\ 
\raggedright ACM Conference, ISSN XXXX-XXXX. \\
\href{https://doi.org/\vldbdoi}{doi:\vldbdoi} \\
}\addtocounter{footnote}{-1}\endgroup

\section{Introduction}
\label{sec:intro}

Time Series Generation (TSG) has become a cornerstone technique for tasks such as data augmentation \cite{bao2024towards, t-cgan}, anomaly detection \cite{cad, wang2021tsagen}, privacy preservation \cite{pategan, tian2024reliable}, and domain adaptation \cite{sasa, li2022towards}.
The core objective of TSG is to produce synthetic sequences that faithfully replicate the temporal dependencies and cross-dimensional correlations of real-world time series.
Recent years have seen rapid advances in TSG models, supported by benchmarking frameworks like TSGBench \cite{Ang2023TSGBench, ang2024tsgassist}. 
However, existing efforts largely target generic domains (e.g., healthcare, traffic, and industrial signals) and overlook the distinct behaviors and structural complexities present in financial markets.

Cryptocurrencies have recently emerged as a major financial asset class, with the market reaching an estimated value of \$4 trillion by May 2025 \cite{reuters_crypto4trillion}. 
Unlike traditional financial instruments, cryptocurrency markets are characterized by high-frequency global activity, speculative dynamics, and unique microstructures shaped by their decentralized nature. Notable features include:
\begin{itemize}[left=2pt]
  \item \textbf{24/7 Operation:} Trading occurs continuously without centralized market hours or scheduled closures.

  \item \textbf{Lack of Intrinsic Valuation:} With no fundamental disclosures, most tokens rely solely on price and volume for analysis.

  \item \textbf{Extreme Volatility:} Prices are highly sensitive to news, liquidity imbalances, and speculative trading, often without underlying economic anchors.

  \item \textbf{Irregular Liquidity:} Many tokens suffer from inconsistent liquidity, exacerbating price impact and risk exposure.
\end{itemize}

These characteristics defy assumptions in existing financial time series benchmarks \cite{Hu2025FinTSB, wang2025fintsbridge, qiu2024tfb}, which often rely on regular trading hours, stable volatility, or intrinsic valuation anchors. 
This underscores the need for a dedicated benchmark that captures the unique dynamics of crypto markets. 
Accurately modeling and evaluating crypto time series is both methodologically challenging and essential for building robust trading strategies and risk controls.

\subsection{Motivations}
\label{sec:intro:motivation}

Existing benchmarks designed for financial time series \cite{Hu2025FinTSB, wang2025fintsbridge} primarily target traditional financial markets and predominantly emphasize forecasting tasks. Although they have significantly contributed to benchmarking practices, they exhibit critical limitations (\textbf{L1--L3}) when applied to cryptocurrency markets:

\paragraph{L1: The limited domain generality hinders evaluation under cryptocurrency's round-the-clock volatility}
\textsf{TSGBench} \cite{Ang2023TSGBench} includes a diverse collection of real-world time series; however, its financial data coverage is notably limited, comprising only a single stock dataset and an exchange dataset. 
Similarly, benchmarks like \textsf{FinTSB} \cite{Hu2025FinTSB} predominantly feature stock indices (e.g., CSI300), which inherently exhibit lower volatility compared to cryptocurrencies. 
These benchmarks overlook cryptocurrency data spanning numerous tokens and trading pairs, thus lacking explicit support for cryptocurrency data, despite its growing significance and unique market characteristics.

\paragraph{L2: The narrow task scope prioritizes forecasting, leaving crypto-specific generation and trading tasks untested}
Existing financial time series benchmarks primarily target classification and forecasting tasks \cite{Ang2023TSGBench, Hu2025FinTSB, wang2025fintsbridge, wang2024deep}. 
For instance, \textsf{FinTSB} and \textsf{FinTSBridge} focus almost entirely on forecasting, overlooking trading-centric tasks such as arbitrage and strategy evaluation, which are crucial for real-world applications. 
Moreover, current studies seldom explore TSG methods in cryptocurrency contexts, leaving a gap between synthetic generation and actionable financial insights.

\paragraph{L3: The absence of crypto-specific evaluation obscures the models' real trading utility}
Existing benchmarks usually omit crucial financial evaluation metrics essential for a realistic assessment of trading strategies and market-informed decision-making. 
For example, TSGBench \cite{Ang2023TSGBench} emphasizes general fidelity but does not evaluate the practical viability of synthetic data in financial domains. 
While \textsf{FinTSB} \cite{Hu2025FinTSB} introduces some realistic metrics, it remains anchored to traditional stock market conventions such as scheduled market closures and moderate volatility.
Thus, they fail to capture cryptocurrency-specific phenomena such as extreme price volatility, uninterrupted trading dynamics, and risk profiles.

\subsection{Our Contributions}
\label{sec:intro:contributions}

To address these limitations, we introduce \textsf{CTBench}, an open-source benchmark designed explicitly for rigorous evaluation of synthetic TSG methods within the cryptocurrency domain. 
By providing a structured and crypto-centric framework, \textsf{CTBench} significantly advances existing evaluation standards through the following key contributions (\textbf{C1--C4}):

\begin{figure}[t]
  \centering
  \includegraphics[width=0.99\columnwidth]{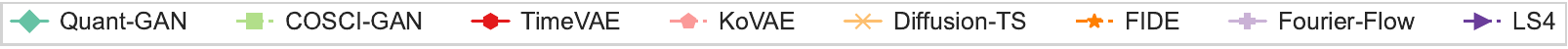} \\
  \includegraphics[width=0.49\columnwidth]{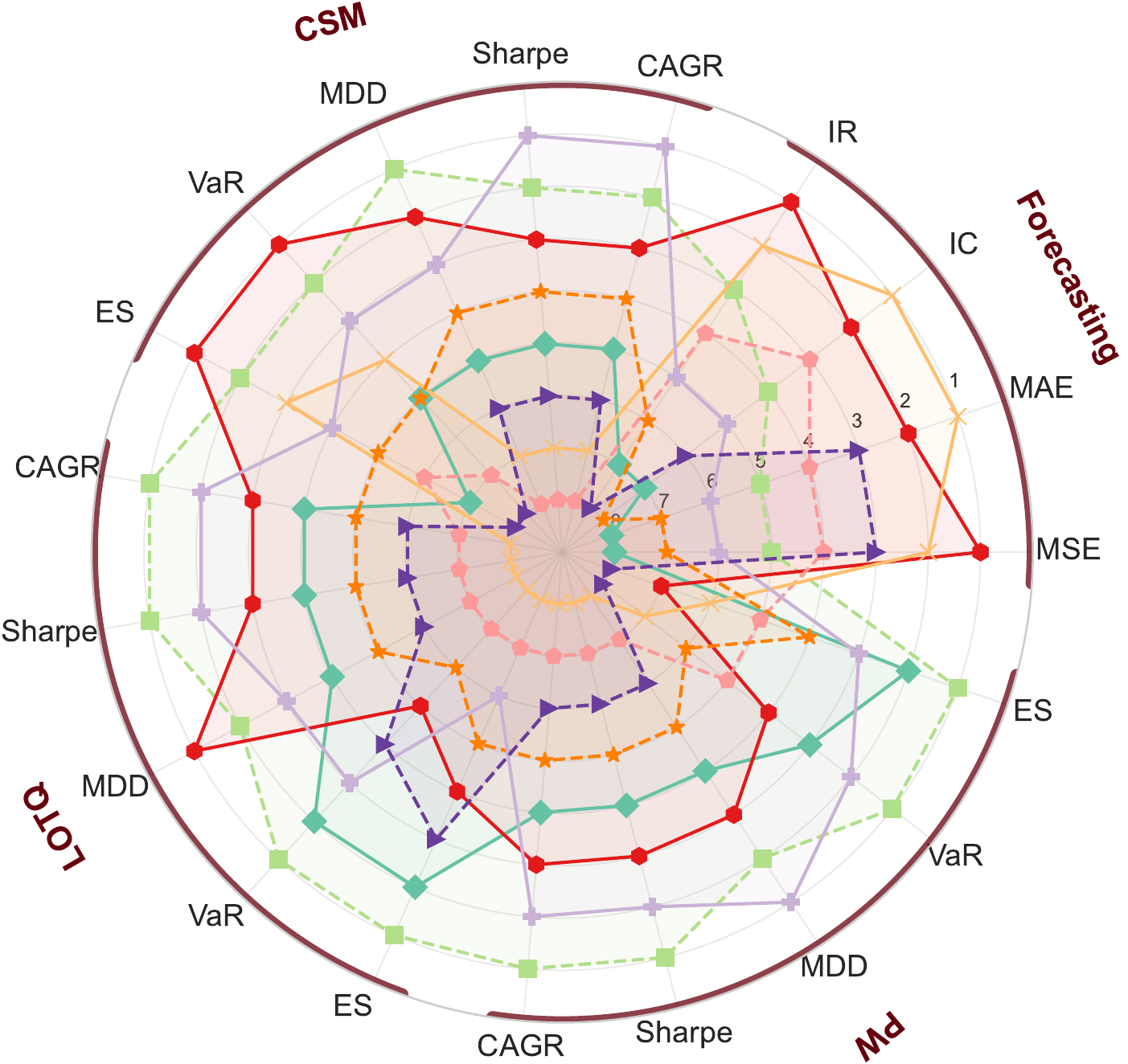}
  \includegraphics[width=0.49\columnwidth]{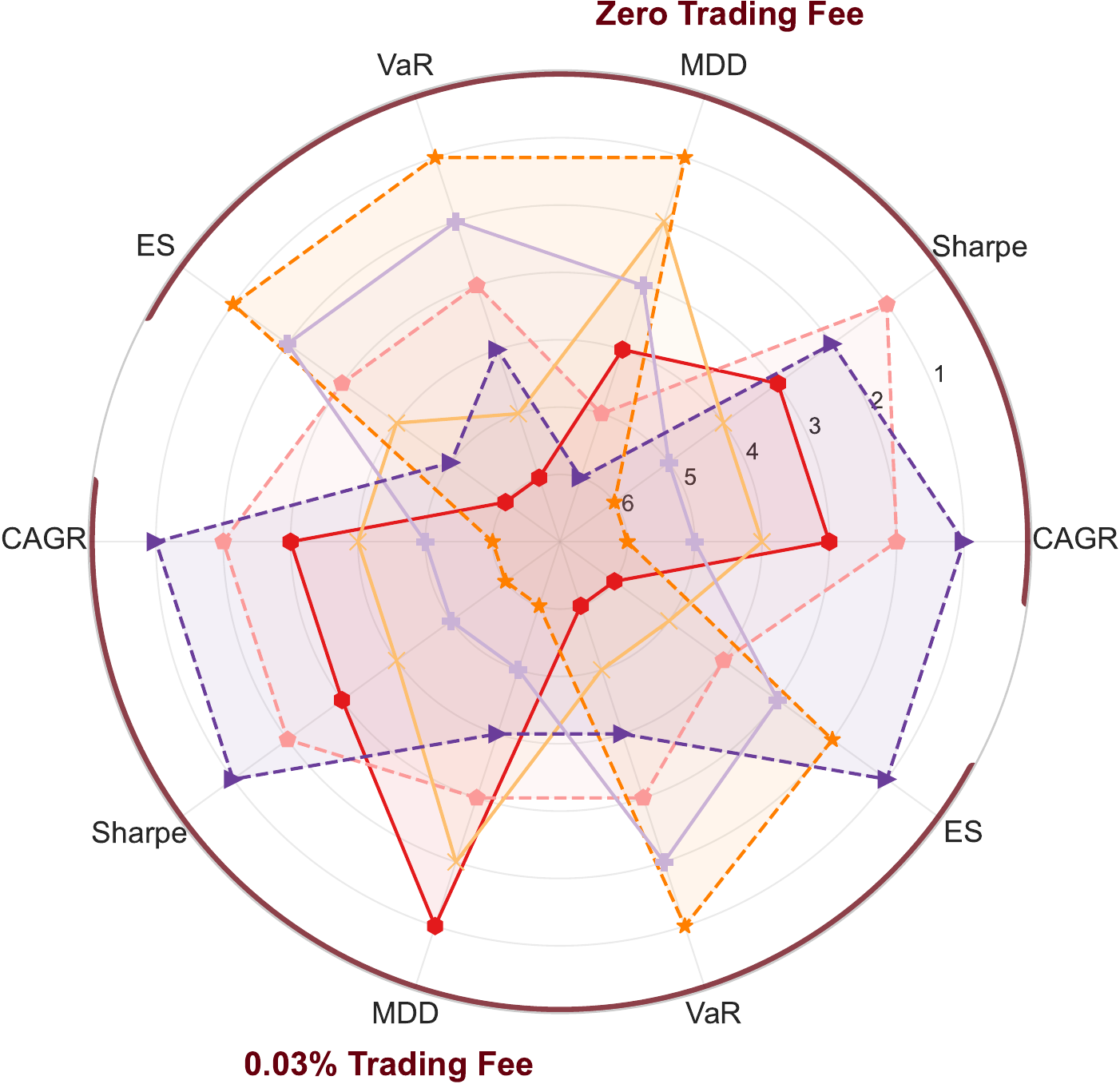}
  \vspace{-0.5em}
  \caption{TSG model rankings on the Predictive Utility (left) and Statistical Arbitrage (right) tasks from 2021 to 2024.}
  \label{fig:overall_ranking}
  \vspace{-0.5em}
\end{figure}

\paragraph{C1: We provide a crypto-centric time series dataset for high-volatility evaluation}
For \textbf{L1}, we present a meticulously curated, publicly available cryptocurrency dataset collected from major global exchanges (\sec{\ref{sec:benchmark:data}}). 
The data undergoes a standardized preprocessing pipeline with configurable options and feature selections tailored to the unique dynamics of crypto markets.
This careful curation ensures high-quality, analysis-ready data that faithfully captures the complexity and volatility inherent to cryptocurrency trading environments.

\paragraph{C2: We design dual-task benchmarks linking TSG to cryptocurrency forecasting and arbitrage}
To address \textbf{L2} and bridge TSG with practical financial applications, \textsf{CTBench} introduces a dual-task evaluation framework that assesses both predictive fidelity and tradability of generated data (\sec{\ref{sec:benchmark:task}}). 
Unlike prior benchmarks focused on reconstruction or statistical similarity, our framework evaluates whether synthetic data can drive actionable outcomes in real-world financial settings:
\begin{itemize}[left=2pt]
  \item \textbf{Predictive Utility Task:} 
  Building on the model-based evaluation paradigm \cite{Ang2023TSGBench}, we design a forecasting-centric task tailored to the nuances of cryptocurrency markets. 
  Synthetic series are used to train forecasting models, which are then evaluated on real market data. 
  Performance reflects how well the synthetic data preserves temporal and cross-sectional structures critical for downstream prediction.
  
  \item \textbf{Statistical Arbitrage Task:} 
  This task examines whether TSG models can capture tradable structures by reconstructing historical returns.
  The residuals from the reconstruction are modeled as mean-reverting signals and fed into statistical arbitrage strategies. 
  Financial metrics on profitability and risk profiles evaluate whether the synthetic data reveal useful trading signals suitable for profitable trading.
\end{itemize}

\paragraph{C3: We construct a holistic financial evaluation measure suite tailored to crypto trading realities}
Regarding \textbf{L3}, to facilitate thorough and realistic financial analyses, \textsf{CTBench} incorporates a comprehensive suite of evaluation measures over diverse trading strategies (\sec{\ref{sec:benchmark:strategy}}) spanning forecasting performance, rank-based predictive measures, key trading performance indicators, and critical risk assessment metrics (\sec{\ref{sec:benchmark:eval}}).

\paragraph{C4: We perform systematic evaluations and distill actionable insights for TSG methods in crypto domains} 
We conduct extensive evaluations across various TSG models (\sec{\ref{sec:benchmark:methods}}).
Through detailed results and ranking analysis, we deliver valuable insights into both the synthetic data generation fidelity and the practical performance of generated data in realistic trading contexts (\sec{\ref{sec:expt}}). 
Figure~\ref{fig:overall_ranking} visualizes the aggregate rankings across two tasks, with metrics arranged radially and performance averaged over strategies and fee scenarios. The results highlight no universally dominant model, revealing trade-offs between fidelity, tradability, and robustness. 
\textsf{CTBench} thus enables informed method selection and strategic refinement tailored to cryptocurrency trading applications.
\section{Preliminaries}
\label{sec:prelim}

\subsection{Problem Definition}
\label{sec:prelim:definition}

Let $\bm{R} \in\mathbb{R}^{n\times l}$ denote the log-return matrix, where $n$ is the number of tradable crypto-assets and $l$ is the number of hourly observations of returns.
At time $t \geq 1$, the log-return vector across all assets is $\bm{r}_t = [r_{1,t}, \cdots, r_{n,t}] \in \mathbb{R}^n$, with each element defined as $r_{i,t} = \log \frac{p_{i,t}}{p_{i,t-1}}$, where $p_{i,t}$ is the price of asset $i$ at hour $t$.

To simulate real-world backtesting, we adopt a rolling-window approach.
Given a training window size $w$ and a test step $s$, we define split offsets $\tau \in \mathcal{O} = \{w, w + s, \cdots, w + (k-1) \times s\}$ with $k = \floor{\tfrac{l-w}{s}}$.
Each split produces a training and test set: 
\begin{displaymath}
  \bm{R}_{\text{train}}^{(\tau)} = [\bm{r}_{\tau-w+1}, \cdots, \bm{r}_{\tau}],~\bm{R}_{\text{test}}^{(\tau)} = [\bm{r}_{\tau+1}, \cdots, \bm{r}_{\tau+s}].
\end{displaymath}

For each split, a Time Series Generation (TSG) model $\bm{g}^{(\tau)}$ is trained on $\bm{R}_{\text{train}}^{(\tau)}$ and evaluated in two modes:
\begin{itemize}[nolistsep,left=2pt] %
  \item \textbf{Generation Mode}: Samples synthetic sequences from Gaussian noise:
  \begin{displaymath}
    \bm{R}_{\text{gen}} = \bm{g}^{(\tau)}(\bm{z}),~\bm{z} \sim \mathcal{N}(\bm{0},\,\bm{I}).
  \end{displaymath}
    
  \item \textbf{Reconstruction Mode}: Reconstructs the train and test set from itself, respectively:
  \begin{displaymath}
    \hat{\bm{R}}_{\text{train}} = \bm{g}^{(\tau)}(\bm{R}_{\text{train}}^{(\tau)}),~\hat{\bm{R}}_{\text{test}} = \bm{g}^{(\tau)}(\bm{R}_{\text{test}}^{(\tau)}).
  \end{displaymath}
\end{itemize}

We also define a basic portfolio simulation setup. 
Starting from an initial capital $V_{0} > 0$, the strategy allocates funds at each hour $t \in \{1, \cdots, l \}$ using a weight vector 
\begin{displaymath}
\bm{\eta}_{t}=[\eta_{1,t}, \cdots, \eta_{n,t}] \in \mathbb{R}^n, \eta_{i,t}=1, 
\end{displaymath}
where $\eta_{i,t}$ denotes the portfolio fraction assigned to asset $i$.
The portfolio evolves as $V_{t}=V_{t-1} \times (\bm{\eta}_{t} \cdot \bm{r}_{t})$,
and the hourly profit–and–loss is defined as $\Delta V_{t} = V_{t}-V_{t-1}$. 
A summary of frequently used notations is provided in Table~\ref{tab:notations}.

\subsection{Scope Illustration}
\label{sec:prelim:scope}

To maintain a clear focus and practical relevance, \textsf{CTBench} explicitly defines its scope across four dimensions: datasets, trading strategies, evaluation measures, and TSG models.

\paragraph{Scope of Datasets}
\textsf{CTBench} is restricted to cryptocurrency markets due to their unique properties, such as 24/7 trading, high volatility, and fragmented liquidity. 
We use only raw time series inputs (i.e., returns and volumes), excluding side-channel information (e.g., order books, blockchain logs, or news). 
This isolates core generative capabilities without reliance on auxiliary signals.
We employ only well-established financial features (e.g., Alpha101 \cite{kakushadze2016101}) to ensure compatibility with real-world quantitative trading while minimizing noise from complex feature engineering.

\paragraph{Scope of Trading Strategies}
To capture diverse trading behaviors, we benchmark TSG models across three canonical strategies, ranging from rank-based to magnitude-sensitive and from directional to market-neutral setups.
This ensures a holistic evaluation of whether synthetic data generalizes across real-world trading paradigms or merely overfits to specific signal patterns.

\begin{table}[t]
\centering
\small
\caption{List of frequently used notations.}
\vspace{-1.0em}
\label{tab:notations}
\resizebox{\columnwidth}{!}{%
\begin{tabular}{ll}
    \toprule
    \rowcolor[HTML]{FFF2CC}
    \textbf{Symbol} & \textbf{Description} \\
    \midrule
    $\bm{R} \in \mathbb{R}^{n \times l}$ & Log-return matrix with $n$ assets and $l$ hourly observations \\
    $\bm{r}_t = [r_{i,t}] \in \mathbb{R}^n$ & Log-return vector of time $t$ of all $n$ assets \\
    
    $w, s, k, \tau$ & Training window size, test step, \# splits, split offset \\
    $\bm{R}_{\text{train}},\bm{R}_{\text{test}}$ & Training data of returns, test data of returns \\
    
    $\bm{g}$ & Time series generation (TSG) model \\
    $\bm{R}_{\text{gen}}$, $\hat{\bm{R}}_{\text{train}},\hat{\bm{R}}_{\text{test}}$ & Generated time series, reconstruction of train and test sets \\
    
    $\bm{\eta}_t = [\eta_{i,t}] \in \mathbb{R}^n$ & Portfolio weight vector at hour $t$\\    
    $V_0$, $V_t$, and $\Delta V_t$ & Initial capital, portfolio equity, and profit‐and‐loss at hour $t$ \\

    $O, H, L, C$ & Open, High, Low, and Close (OHLC) prices \\
    $\bm{D} = [\bm{x}_{i,t}]$ & Data tensor \\
    $\bm{\Phi} = \{ \bm{\phi}_j \}_{j=1}^d$ & A feature set $\bm{\Phi}$ with $d$ feature mapping function $\bm{\phi}_j$ \\
    \bottomrule
\end{tabular}}
\end{table}
\setlength{\textfloatsep}{1.5em}

\paragraph{Scope of Evaluation Measures}
Our benchmark incorporates a curated set of evaluation measures widely recognized in financial TSG research \cite{Ang2023TSGBench, quant-gan}, ensuring a holistic assessment of statistical fidelity and financial utility. 
We have excluded metrics with limited practical relevance or interpretability to maintain a focused and meaningful evaluation framework for the crypto domain.

\paragraph{Scope of TSG Models}
We select TSG models capable of handling multivariate inputs typical in crypto markets, encompassing both general-purpose and finance-specific architectures. 
Our selection spans five major model families (i.e., GAN, VAE, diffusion, flow, and mixed-type), ensuring diverse architectural coverage. 
Models limited to narrow domains or requiring specialized data are excluded to preserve general applicability. 
All models are trained under a unified protocol without excessive hyperparameter tuning to ensure fair benchmarking and reflect practical deployment constraints.

\section{CTBench}
\label{sec:benchmark}

\begin{figure*}[t]
  \centering
  \includegraphics[width=0.99\textwidth]{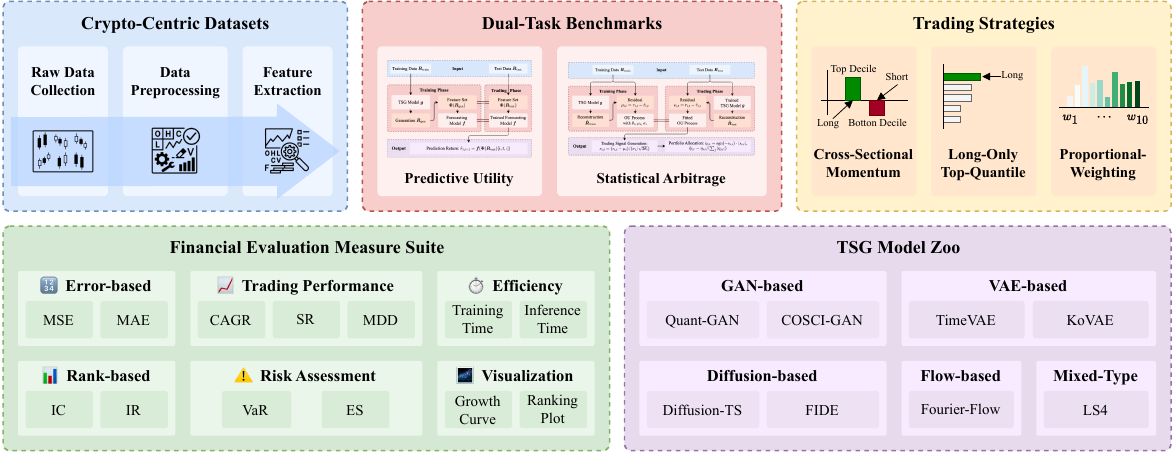}
  \vspace{-0.75em}
  \caption{Overall architecture of \textsf{CTBench}.}
  \label{fig:ctbench}
  \vspace{-0.75em}
\end{figure*}

We present \textsf{CTBench}, a comprehensive benchmark customized for evaluating TSG models in the context of cryptocurrency markets.
As illustrated in Figure~\ref{fig:ctbench}, \textsf{CTBench} integrates five key modules that provide a rigorous and versatile evaluation framework:
\begin{enumerate}[nolistsep,leftmargin=16pt]
  \item \textbf{Crypto-Centric Datasets (\sec{\ref{sec:benchmark:data}}):} 
  Hourly 24/7 OHLC data from 452 cryptocurrencies, curated and processed via a standardized pipeline for consistency and reliability.

  \item \textbf{Dual-Task Benchmarks (\sec{\ref{sec:benchmark:task}}):} 
  Two complementary tasks--Predictive Utility and Statistical Arbitrage--evaluate both predictive fidelity and practical utility by testing signal preservation and tradability.

  \item \textbf{Trading Strategies (\sec{\ref{sec:benchmark:strategy}}):} 
  Three diverse strategies stress-test how well synthetic data supports various trading styles, reducing the risk of model overfitting.
  
  \item \textbf{Financial Evaluation Measure Suite (\sec{\ref{sec:benchmark:eval}}):} 
  Thirteen metrics encompassing prediction errors, rank fidelity, trading performance, risk assessment, and efficiency offer a holistic view of statistical quality and economic utility.

  \item \textbf{TSG Model Zoo (\sec{\ref{sec:benchmark:methods}}):} 
  Eight representative TSG models spanning VAEs, GANs, diffusion, flow-based, and mixed-type approaches enable fair, architecture-agnostic comparisons.
\end{enumerate}

\subsection{Crypto-Centric Datasets}
\label{sec:benchmark:data}

\paragraph{Data Overview and Preprocessing}
We construct our datasets using historical hourly data for all spot trading pairs listed on the Binance exchange \cite{binance_data}. 
The data spans from January 2020 to December 2024, ensuring coverage across diverse market regimes, including bull runs, crashes, and consolidation phases. 
To guarantee high data quality, we filter out assets with missing observations and restrict our dataset to pairs traded against USDT (Tether). 
The resulting dataset comprises 452 unique cryptocurrencies, offering a robust foundation for TSG benchmarking in crypto markets.

Formally, let $n$ denote the number of tradable crypto assets and $(l+1)$ the number of hourly observations after data filtering.
We index assets by $1 \leq i \leq n$ and timestamps by $0 \leq t \leq l$. 
For each asset and timestamp pair $(i,t)$, we record the five standard fields:
\begin{displaymath}
  \bm{x}_{i,t} = [O_{i,t}, H_{i,t}, L_{i,t}, C_{i,t}] \in \mathbb{R}^{4},
\end{displaymath}
where $O$, $H$, $L$, and $C$ represent the \textbf{Open}, \textbf{High}, \textbf{Low}, and \textbf{Close} prices (quoted in USDT). 
Stacking these observations yields the data tensor:
\begin{displaymath}
  \bm{D} = [\bm{x}_{i,t}] \in \mathbb{R}^{n \times (l+1) \times 4}.
\end{displaymath}

In this work, we focus primarily on the close prices and define hourly log-returns as: $r_{i,t} = \log \tfrac{C_{i,t}}{C_{i,t-1}}$, where $t \in \{1, \cdots, l\}$.
The complete return matrix is $\bm{R} = [r_{i,t}] \in \mathbb{R}^{n \times l}$.

\paragraph{Feature Extraction}
To capture essential market dynamics, we engineer a diverse set of $d$ scalar features commonly used in quantitative trading. 
These include Alpha101 factors \cite{Kakushadze2016FormulaicAlphas} and traditional technical indicators such as Bollinger Bands, RSI, and moving averages. 
Such features encode signals related to momentum, mean-reversion, volatility, and other short-term market behaviors, widely leveraged in quantitative finance research \cite{sun2023reinforcement, zhualphaqcm, zhang2020doubleensemble, tsai2010combining}.

Applying the same feature-extraction pipeline to both real and synthetic datasets allows us to isolate and rigorously test the TSG models' capacity to replicate the statistical and structural properties vital for downstream tasks.
Formally, let $\bm{\Phi}=\{ \bm{\phi}_j \}_{j=1}^{d}$ be the feature set, where each $\bm{\phi}_j: \mathbb{R}^{n \times l} \to \mathbb{R}^{n \times l}$ acts on the return matrix $\bm{R}$. 
Applying $\bm{\Phi}$ yields the feature tensor with shape $ \mathbb{R}^{n \times l \times d}$.

\paragraph{Data Descriptive Statistics}
Understanding the statistical profile of crypto returns is essential for designing effective TSG benchmarks. 
We analyze the distribution of log-returns to identify deviations from normality, such as skewness and kurtosis--stylized facts well documented in financial time series.
Cryptocurrencies, in particular, often exhibit \textbf{fat-tailed distributions}, indicating elevated probability of extreme price movements. 

\begin{figure}[ht]
  \centering
  \includegraphics[width=0.48\columnwidth]{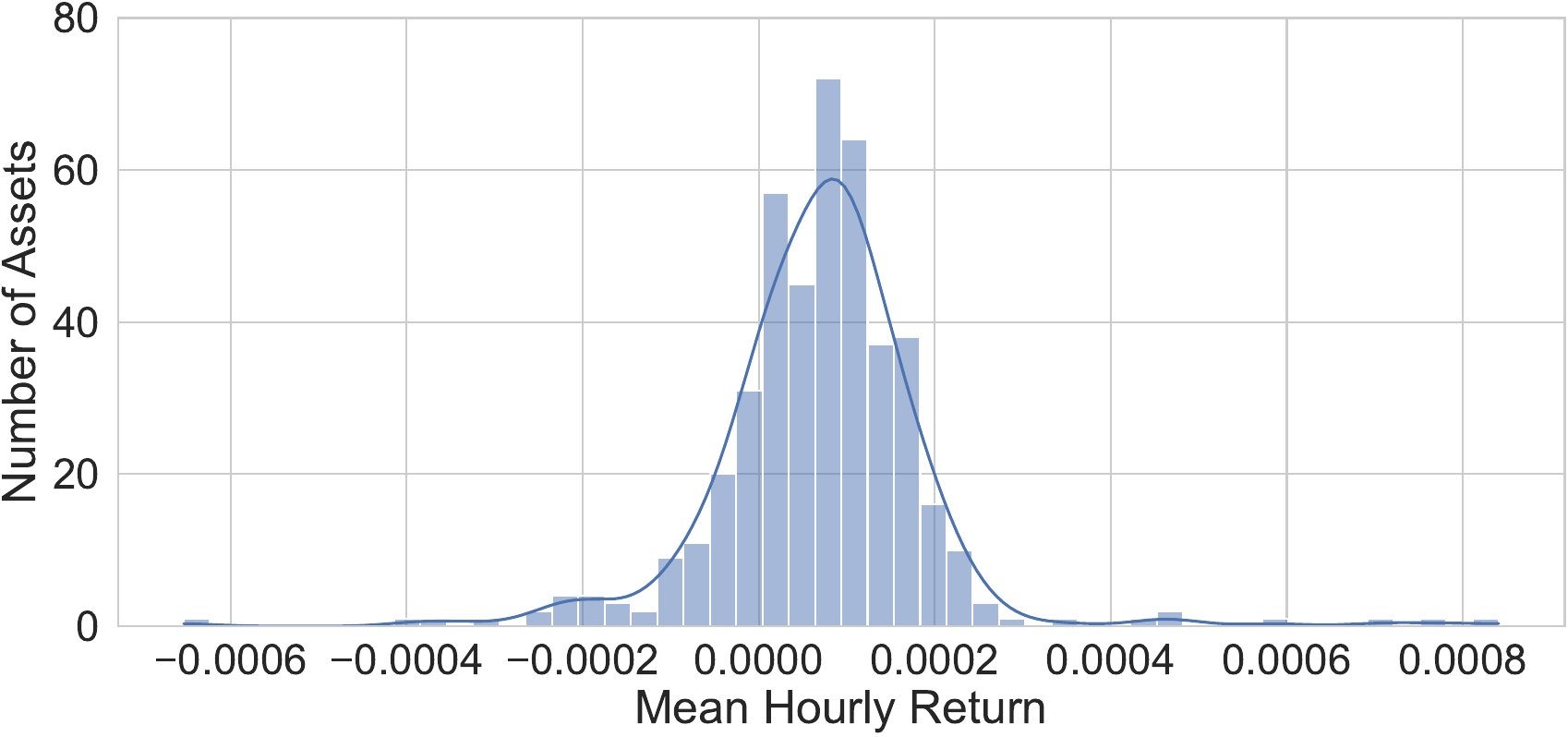}
  \hfill
  \includegraphics[width=0.48\columnwidth]{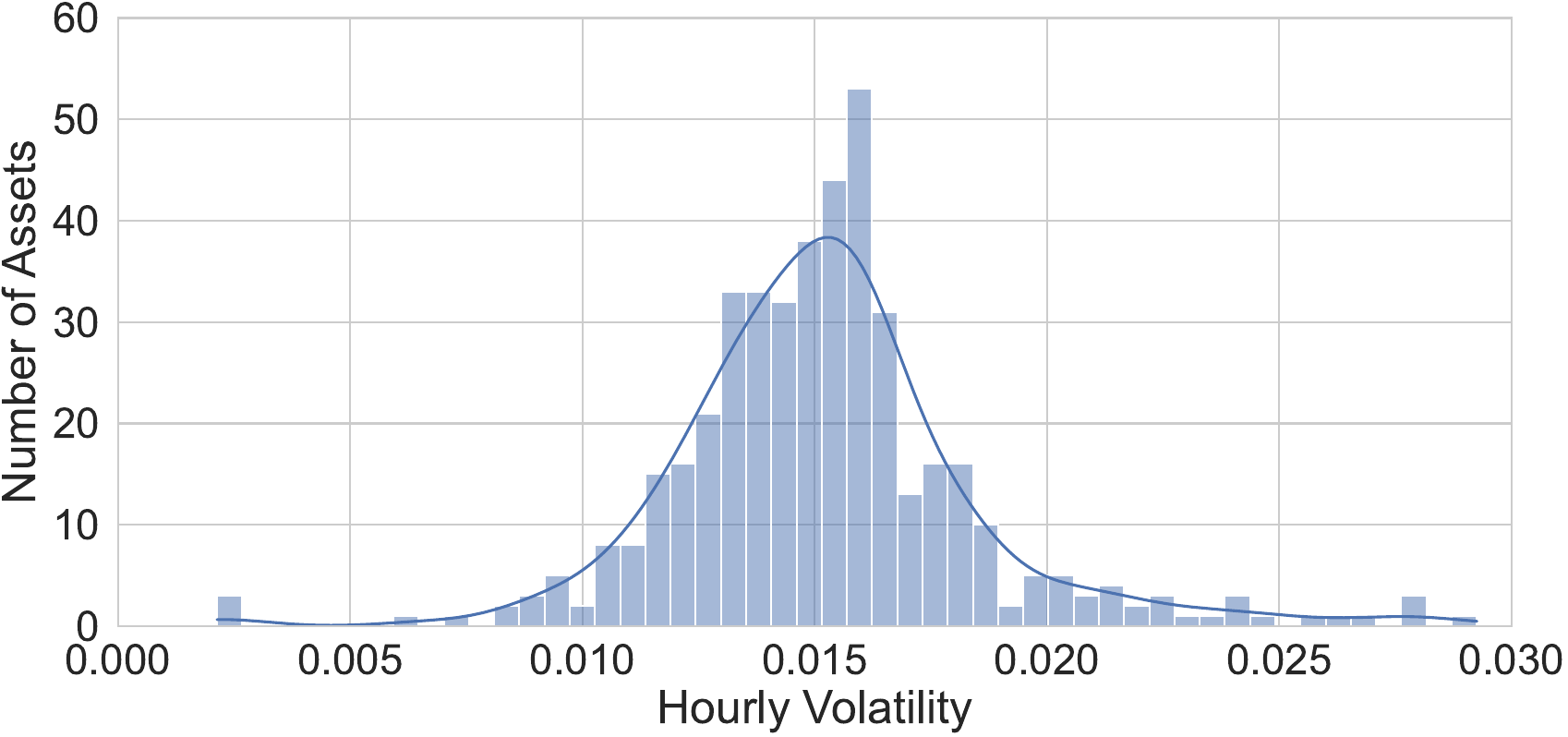}
  \vspace{-1.0em}
  \caption{Histograms of the mean hourly log-return (\%) (left) and mean hourly volatility (\%) (right).}
  \label{fig:histo-mean-return-and-vol}
  \vspace{-0.5em}
\end{figure}

Figure~\ref{fig:histo-mean-return-and-vol} presents histograms of the mean hourly log-return and mean hourly volatility (standard deviation of log-returns) across all 452 cryptocurrencies.
The mean hourly returns are centered around zero but show a slight right skew, suggesting modestly positive drift in most assets. 
In contrast, the mean hourly volatility exhibits a long right tail, indicating that while many assets trade with low volatility, a notable subset experiences highly volatile price swings.

\begin{figure*}[t]
  \centering
  \includegraphics[width=0.99\textwidth]{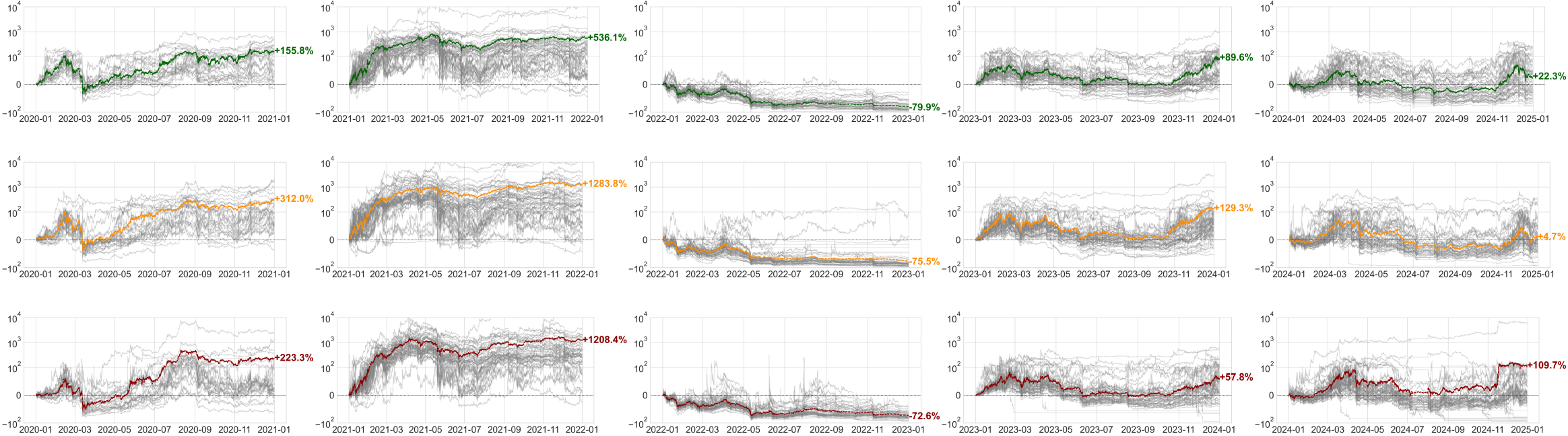}
  \vspace{-1.0em}
  \caption{Line plots of closing returns for representative cryptocurrencies, with large-cap examples (top row), mid-cap examples (middle row), and small-cap examples (bottom row), displayed annually from 2020 to 2024.}
  \label{fig:price_line_plot}
  \vspace{-0.5em}
\end{figure*}

To visualize market dynamics over time, we categorize cryptocurrencies into large-, mid-, and small-cap groups and plot representative closing prices annually from 2020 to 2024 in Figure~\ref{fig:price_line_plot}.
These trajectories highlight significant market regimes, including the bull runs of 2020--2021, sharp corrections in 2022, and subsequent periods of recovery or consolidation. 
Notably, mid- and small-cap assets often display greater volatility and sharper price swings than their large-cap counterparts.

\begin{figure}[t] 
  \centering
  \includegraphics[width=0.48\columnwidth]{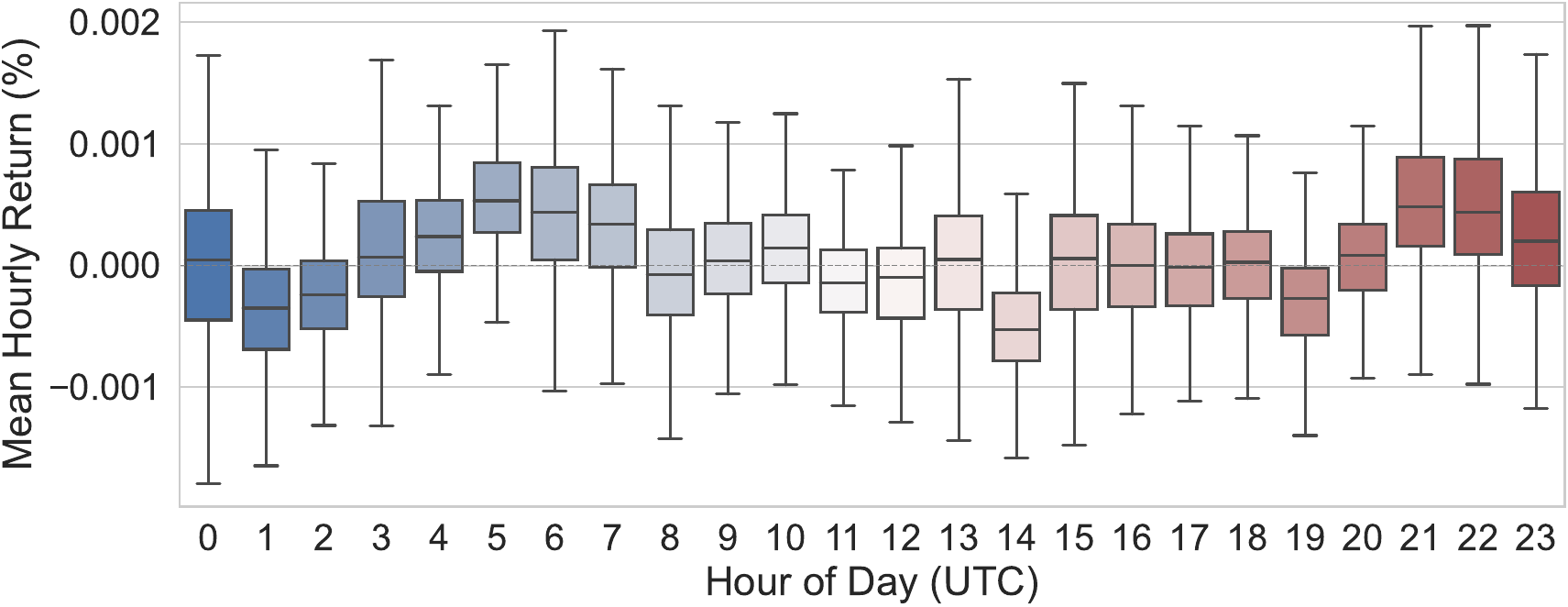}
  \hfill
  \includegraphics[width=0.48\columnwidth]{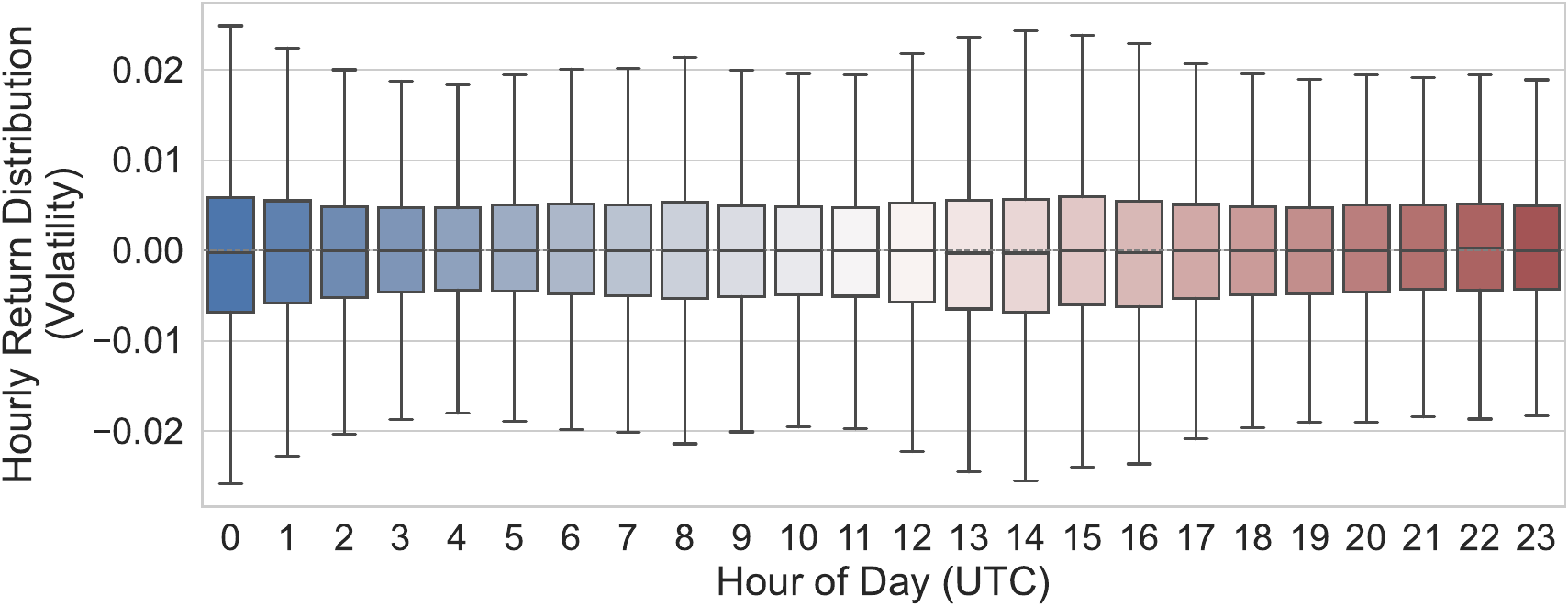}
  \vspace{-1.0em}
  \caption{The mean hourly log-return (\%) (left) and mean hourly volatility (\%) (right) by hour of day (UTC).}
  \label{fig:hour}
  \vspace{-0.5em}
\end{figure}

Given that cryptocurrency markets operate 24/7, intraday patterns provide valuable insights into market microstructure. 
Figure~\ref{fig:hour} depicts the mean hourly log-return and volatility by time of day. 
We observe return peaks around early morning (5--7 AM) and late evening (9--11 PM), reflecting heightened trading during transitions between major global financial centers. 
Volatility peaks notably around midnight and during overlapping trading hours between US and Europe (12--5 PM), suggesting periods of intensified market activity driven by global participation and algorithmic strategies.

\paragraph{Discussions}
Our analysis reveals several critical insights shaping the design of \textsf{CTBench}:
\begin{itemize}[nolistsep,left=2pt] %
  \item \textbf{Complex Market Dynamics:}
  Crypto markets exhibit high-frequency, high-dimensional behaviors with distinct volatility profiles, intraday cycles, and regime shifts. These factors necessitate benchmarks tailored for crypto time series.

  \item \textbf{Benchmark Task Design:}
  Given the data's complexity, evaluation tasks must probe whether synthetic data preserves predictive structures critical for practical applications such as forecasting and statistical arbitrage.

  \item \textbf{TSG Model Requirements:}
  Capturing the intricate temporal and cross-sectional dependencies of crypto markets demands advanced TSG architectures capable of modeling both short-term fluctuations and long-term trends.

  \item \textbf{Evaluation Metrics:}
  Assessing TSG performance in crypto markets requires multifaceted metrics that go beyond statistical fidelity to capture financial viability and risk sensitivity.
\end{itemize}

Collectively, these insights underscore the need for crypto-specific benchmarks like \textsf{CTBench} to advance the evaluation and development of TSG models for this rapidly evolving domain.

\subsection{Dual-Task Benchmarks}
\label{sec:benchmark:task}

\begin{figure*}[t]
\centering
\subfigure[Predictive Utility task.]{
  \label{fig:ctbench_dall_task:gfb}
  \includegraphics[width=0.41\textwidth]{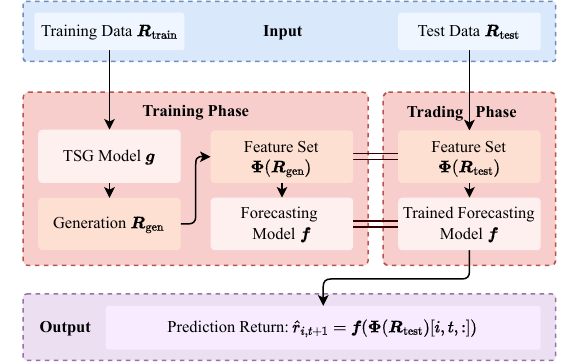}}
\subfigure[Statistical Arbitrage task.]{
  \label{fig:ctbench_dall_task:rtb}
  \includegraphics[width=0.56\textwidth]{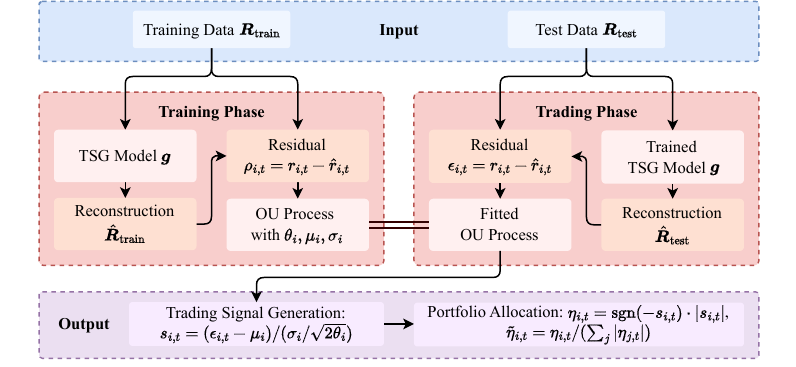}}
\vspace{-1.25em}
\caption{Architectures of dual-task benchmarks.}
\label{fig:ctbench_dall_task}
\vspace{-1.0em}
\end{figure*}

To bridge synthetic TSG with practical financial use, \textsf{CTBench} introduces dual-task benchmarks assessing both statistical similarity and the functional realism and trading utility of synthetic data. 
As illustrated in Figure~\ref{fig:ctbench_dall_task}, these tasks probe complementary aspects of TSG models: generation quality through predictive utility and reconstruction fidelity via tradable residual signals.

\subsubsection{Predictive Utility Task}
\label{sec:benchmark:task:forecasting}

This task evaluates whether synthetic data generated by TSG models can effectively train \emph{forecasting models} that perform well on real-world market data.
Different from likelihood metrics or two-sample statistical tests, this task measures economic value: synthetic data are judged by the trading performance they enable. 
Figure~\ref{fig:ctbench_dall_task:gfb} depicts the workflow.

\paragraph{Training Phase}
Let $\bm{R}_{\text{train}}^{(\tau)} = [\bm{r}_{\tau - w + 1}, \cdots, \bm{r}_{\tau}] \in \mathbb{R}^{n \times w}$ denote the real log-return matrix for a split offset $\tau$ with length $w=500\times24$ hours.  
A TSG model $\bm{g}$ is trained on $\bm{R}_{\text{train}}^{(\tau)}$ to capture both temporal dependencies and cross-sectional relationships.
From this trained model, we sample synthetic returns:
\begin{displaymath}
  \bm{R}_{\text{gen}} = \bm{g}(\bm{z}), \bm{z} \sim \mathcal{N}(\bm{0},\,\bm{I})
\end{displaymath}

Next, features are extracted from $\bm{R}_{\text{gen}}$ via the pipeline: $\bm{\Phi}(\bm{R}_{\text{gen}})\in\mathbb{R}^{n \times s \times d}$.
A forecasting model $\bm{f}: \mathbb{R}^{d} \to \mathbb{R}$ then predicts the next-hour return: 
\begin{displaymath}
  \hat{r}_{i,t+1}=\bm{f}(\bm{\Phi}(\bm{R}_{\text{gen}})[i,t,:]).
\end{displaymath}

We use XGBoost \cite{chen2016xgboost} as the forecasting model, chosen for its balance of robustness, interpretability, and minimal hyperparameter tuning \cite{vancsura2025navigating, liu2021multi, yun2021prediction}, ensuring that benchmark results primarily reflect the quality of the generated data rather than model capacity.

\paragraph{Trading Phase}
The trained forecaster is then applied to a test period of length $ s = 30 \times 24$ hours.
For each hour $t$ and asset $i$, we predict $\hat r_{i,t+1} = \bm{f}(\bm{\Phi}(\bm{R}_{\text{test}})[i,t,:])$, rank the vector $\hat{\bm{r}}_{t+1}=[\hat r_{i,t+1}]_{i=1}^{n}$, and construct a dollar-neutral portfolio by \textbf{longing} the top half of assets (highest $\hat{r}_{i, t+1}$) and \textbf{shorting} the bottom half (lowest $\hat{r}_{i,t+1}$).
This portfolio is rebalanced hourly over the test window, maintaining balanced long and short exposures.

\paragraph{Discussions}
This task reveals how well synthetic data generalizes to real markets, operationalizing the notion of functional realism. 
If $\bm{R}_{\text{gen}}$ preserves the predictive structures of $\bm{R}_{\text{train}}^{(\tau)}$, the realized P\&L $\Delta V_{t}$ will score highly across \textsf{CTBench}'s evaluation suite. 
Thus, synthetic data are valued not merely for statistical closeness to historical distributions but for the economic utility they unlock.
Importantly, every component in Figure~\ref{fig:ctbench_dall_task:gfb} is modular: researchers can substitute alternative TSG models, forecasters (e.g., Transformers), or feature sets, while retaining a unified scoring framework.

\subsubsection{Statistical Arbitrage Task}
\label{sec:benchmark:task:trading}

In contrast to the generation-focused task, the Statistical Arbitrage task assesses a TSG model's ability to \emph{reconstruct} real market dynamics and isolate tradable residual signals. 
Here, the model acts as a ``denoiser,'' stripping away common market components to reveal residuals suitable for statistical arbitrage. 
Figure~\ref{fig:ctbench_dall_task:rtb} summarizes the pipeline.

\paragraph{Training Phase}
The Statistical Arbitrage task typically hinges on pairs or baskets of assets whose spreads revert toward a long-term mean. 
In this task, residuals between real $\bm{R}_{\text{train}}$ and reconstructed returns $\hat{\bm{R}}_{\text{train}}$ are assumed to follow mean-reverting dynamics. 
For asset $i$ and time $t$, we define training residual:
\begin{displaymath}
  {\rho}_{i,t} = r_{i,t} - \hat{r}_{i,t},
\end{displaymath} 
where $r_{i,t} \in \bm{R}_{\text{train}}$ and $\hat{r}_{i,t} \in \hat{\bm{R}}_{\text{train}}$. For each asset $i$, these residuals are then fitted to an Ornstein--Uhlenbeck (OU) process \cite{uhlenbeck1930theory}:
\begin{displaymath}
  d{\rho}_{i,t} = \theta_i(\mu_i-\rho_{i,t}) d t + \sigma_{i} d W_t,
\end{displaymath} 
where $\theta_i > 0$ (mean reversion speed), $\mu_i$ (long-run mean), and $\sigma_i$ (volatility) are estimated per asset, and $d W_t$ is a standard Wiener increment.  
The framework is flexible, supporting alternative processes such as jump-type Lévy processes \cite{fujiwara1985stochastic} or neural SDEs \cite{oh2024stable}.

\paragraph{Trading Phase}
On test data $\bm{R}_{\text{test}}$, the model reconstructs returns $\hat{\bm{R}}_{\text{test}}$, producing test residuals for $r_{i,t} \in \bm{R}_{\text{test}}$ and $\hat{r}_{i,t} \in \hat{\bm{R}}_{\text{test}}$:
\begin{displaymath}
  \epsilon_{i,t} = r_{i,t} - \hat{r}_{i,t}.
\end{displaymath}
Each residual $\epsilon_{i,t}$ is converted to an $s$-score:
\begin{displaymath}
  s_{i,t} = \tfrac{\epsilon_{i,t} - \mu_i}{\sigma_i/\sqrt{2\theta_i}},
\end{displaymath}
quantifying the deviation from equilibrium. 
Trading signals are then derived via:
\begin{itemize}[nolistsep,left=2pt] %
  \item \textbf{Thresholding:} Open or maintain a short position if $s_{i,t}>+\gamma$, a long if $s_{i,t}<-\gamma$, otherwise stay flat, with $\gamma=2$.
  
  \item \textbf{Weight Normalization:} Raw signals $\eta_{i,t} = \operatorname{sgn}(-s_{i,t}) \cdot |s_{i,t}|$ are normalized to $\tilde \eta_{i,t} = \eta_{i,t} / (\sum_j | \eta_{j,t}|)$.
  
  \item \textbf{Execution:} Portfolios are rebalanced hourly based on $\tilde{\eta}_{i,t}$.
\end{itemize}

\paragraph{Discussions}
The Statistical Arbitrage task evaluates whether reconstructed time series reveal stable, mean-reverting residuals suitable for statistical arbitrage, complementing the generation-focused task by addressing market-neutral alpha extraction.
These tasks ensure TSG models are tested not only for statistical fidelity but also for practical effectiveness in real-world crypto trading.

\subsection{Trading Strategies}
\label{sec:benchmark:strategy}

A TSG model that excels under a single trading strategy may offer limited value to practitioners whose trading desks rely on diverse alpha signals.
Thus, \textsf{CTBench} is explicitly designed to be \textbf{strategy-agnostic}, evaluating TSG models across a spectrum of trading paradigms to ensure broad applicability.

Rather than focusing solely on one approach, our benchmark computes consistent profitability and risk metrics (see \sec{\ref{sec:benchmark:eval}}) for the profit-and-loss streams from any back-test.
Applying this evaluation across diverse strategies provides a rigorous stress test, revealing whether a TSG model genuinely captures market microstructure or merely overfits specific trading styles.
We summarize three canonical strategies widely used in crypto trading:
\begin{itemize}[nolistsep,left=2pt] %
  \item \keypoint{S1: Cross-Sectional Momentum (CSM)} takes long positions in the top decile and short positions in the bottom decile of assets ranked by predicted one-hour returns. 
  This probes a model's ability to capture ranking-based alpha signals.

  \item \keypoint{S2: Long-Only Top-Quantile (LOTQ)} equally weights and goes long in the top 20\% of assets based on predicted returns, with all other weights set to zero. This isolates pure directional signals without short exposure.

  \item \keypoint{S3: Proportional-Weighting (PW)} allocates weights proportionally to predicted returns: $\eta_{i,t} = {\hat{r}_{i,t}} / ({\sum_{j=1}^{n} \hat{r}_{j,t}})$, emphasizing the magnitude of forecasted signals rather than merely their ranks.
\end{itemize}

Each strategy exploits different statistical regularities, including level effects, cross-sectional dispersion, and serial correlations, ensuring that no single modeling flaw remains undetected.
They span the primary mandates seen on crypto desks: beta-neutral long–short equity, directional trend capture, and volatility harvesting.
Finally, the \textsf{CTBench} pipeline is fully \textbf{plug-and-play}.
Traders can drop in any proprietary strategies without altering the benchmark code, fostering fair comparison across future studies.

\subsection{Financial Evaluation Measure Suite}
\label{sec:benchmark:eval}

Evaluating TSG models for financial applications demands more than mere statistical similarity; it requires assessing whether synthetic data supports practical trading tasks. 
To this end, \textsf{CTBench} organizes eleven well-established evaluation metrics into five categories, each answering a distinct question practitioners face when considering synthetic data for production.

\paragraph{Error-based Evaluation}
At the most fundamental level, models should accurately predict future asset values. 
Error metrics identify systematic biases or large idiosyncratic deviations that might be masked by portfolio-level metrics.
Given the actual return $r_{i,t}$ and prediction $\hat r_{i,t}$ for asset $i$ and time $t$:
\begin{itemize}[nolistsep,left=2pt] %
  \item \keypoint{E1: Mean Squared Error (MSE)} is defined as:
  \begin{displaymath}
    \text{MSE} = \textstyle \frac{1}{k \cdot s \cdot n} \sum_{\tau \in \mathcal{O}} \sum_{t = 1}^{s} \sum_{i = 1}^{n} {(r_{i,t+\tau} - \hat{r}_{i,t+\tau})}^2 .
  \end{displaymath}

  \item \keypoint{E2: Mean Absolute Error (MAE)} is defined as
  \begin{displaymath}
    \text{MAE} = \textstyle \frac{1}{k \cdot s \cdot n} \sum_{\tau \in \mathcal{O}} \sum_{t = 1}^{s} \sum_{i = 1}^{n} | r_{i,t+\tau} - \hat r_{i,t+\tau} |.
  \end{displaymath}
\end{itemize}
Low values in both metrics reflect strong signal fidelity, while differences help distinguish outliers from widespread minor errors.

\paragraph{Rank-based Evaluation}
In many quantitative trading desks, correctly ranking assets is more crucial than precisely predicting return magnitudes. 
These metrics evaluate whether synthetic data preserves cross-sectional relationships among assets \cite{treynor1973use, richard2000active}.
Given realized returns $\bm{r}_t$ and predictions $\hat{\bm{r}}_t$ for all assets at time $t$:
\begin{itemize}[nolistsep,left=2pt] %
  \item \keypoint{E3: Information Coefficient (IC)} is defined as the average Spearman correlation between predicted and actual rankings, where $\mathrm{IC}_{\tau,t} = \mathrm{SpearmanCorr}(\bm{r}_{t+\tau}, \hat{\bm{r}}_{t+\tau})$. It is computed as:
  \begin{displaymath}
    \mathrm{IC} = \textstyle \frac{1}{k \cdot s} \sum_{\tau \in \mathcal{O}} \sum_{t = 1}^{s} \mathrm{IC}_{\tau,t}.
  \end{displaymath}

  \item \keypoint{E4: Information Ratio (IR)} measures the stability of IC:
  \begin{displaymath}
    \mathrm{IR} = \mathrm{Mean}(\mathrm{IC}_{\tau,t}) / \mathrm{Std} (\mathrm{IC}_{\tau,t}).
  \end{displaymath}
\end{itemize}
A consistently positive IC shows the generator preserves rankings essential for long-short strategies, despite absolute errors.

\begin{table}[t]
\centering
\small
\renewcommand{\arraystretch}{1.08}
\caption{Summary of popular TSG methods with their back-
bone models and financial datasets used.}
\vspace{-1.0em}
\label{tab:methods}
\resizebox{0.99\columnwidth}{!}{
\begin{tabular}{llll} 
    \toprule
    \rowcolor[HTML]{FFF2CC}
    \textbf{Year} & \textbf{Method} & \textbf{Backbone} & \textbf{Financial Datasets Used} \\
    \midrule
    \rowcolor[HTML]{DDEBFF}
    2016  & C-RNN-GAN \cite{crnngan}         & GAN & /      \\
    \rowcolor[HTML]{DDEBFF}
    2017  & RCGAN \cite{rcgan}               & GAN & /      \\
    \rowcolor[HTML]{DDEBFF}
    2018  & T-CGAN \cite{t-cgan}             & GAN & /      \\
    \rowcolor[HTML]{DDEBFF}
    2019  & TimeGAN \cite{timegan}           & GAN & Stocks \\
    \rowcolor[HTML]{DDEBFF}
    2019  & WaveGAN \cite{wavegan}           & GAN & /      \\
    \rowcolor[HTML]{DDEBFF}
    2020  & COT-GAN \cite{cot-gan}           & GAN & / \\
    \rowcolor[HTML]{DDEBFF}
    2020  & DoppelGANger \cite{DoppelGANger} & GAN & / \\
    \rowcolor[HTML]{DDEBFF}
    2020  & Quant-GAN \cite{quant-gan}       & GAN & SPX \\
    \rowcolor[HTML]{DDEBFF}
    2020  & SigCWGAN \cite{c-sig-gan}        & GAN & SPX \& DJI \\
    \rowcolor[HTML]{DDEBFF}
    2020  & TSGAN \cite{cgan}                & GAN & / \\
    \rowcolor[HTML]{DDEBFF}
    2021  & RTSGAN \cite{rtsgan}             & GAN & Stocks \\
    \rowcolor[HTML]{DDEBFF}
    2021  & Sig-WGAN \cite{sig-gan}          & GAN & SPX \& DJI \\
    \rowcolor[HTML]{DDEBFF}
    2021  & TimeGCI \cite{imitation}         & GAN & / \\
    \rowcolor[HTML]{DDEBFF}
    2022  & CEGEN \cite{cegen}               & GAN & Stocks \& Electric Price \\
    \rowcolor[HTML]{DDEBFF}
    2022  & COSCI-GAN \cite{cosci-gan}       & GAN & / \\
    \rowcolor[HTML]{DDEBFF}
    2022  & PSA-GAN \cite{psa-gan}           & GAN & / \\
    \rowcolor[HTML]{DDEBFF}
    2022  & TsT-GAN \cite{tst-gan}           & GAN & Stocks \\
    \rowcolor[HTML]{DDEBFF}
    2022  & TTS-GAN \cite{tts-gan}           & GAN & / \\
    \rowcolor[HTML]{DDEBFF}
    2023  & AEC-GAN \cite{aec-gan}           & GAN & / \\
    \rowcolor[HTML]{DDEBFF}
    2023  & TT-AAE \cite{tt-aae}             & GAN & Stocks \\
    \midrule

    2021  & TimeVAE \cite{timevae}     & VAE   & Stocks \\
    2023  & CRVAE \cite{crvae}         & VAE   & /      \\
    2023  & TimeVQVAE \cite{timevqvae} & VAE   & /      \\
    2024  & KoVAE \cite{kovae}         & VAE   & Stocks \\
    \midrule
    
    \rowcolor[HTML]{DDEBFF} 
    2023  & DiffTime \cite{difftime}         & Diffusion & Stocks \\
    \rowcolor[HTML]{DDEBFF} 
    2023  & TSGM \cite{sgm}                  & Diffusion & Stocks \\
    \rowcolor[HTML]{DDEBFF} 
    2024  & Diffusion-TS \cite{diffusionts}  & Diffusion & Stocks \\
    \rowcolor[HTML]{DDEBFF} 
    2024  & FIDE \cite{galib2024fide}        & Diffusion & Stocks \\
    \rowcolor[HTML]{DDEBFF} 
    2024  & ImagenTime \cite{ImagenTime}     & Diffusion & Stocks \\
    \rowcolor[HTML]{DDEBFF} 
    2024  & SDformer \cite{chen2024sdformer} & Diffusion & Stocks \\
    \rowcolor[HTML]{DDEBFF} 
    2025  & PaD-TS \cite{padts}              & Diffusion & Stocks \\
    \midrule
    
    2020  & CTFP \cite{ctfp}            & Flow  & /      \\
    2021  & Fourier-Flow \cite{fourier} & Flow  & Stocks \\
    2024  & FlowTS \cite{hu2024fm}      & Flow  & Stocks \\
    \midrule
    
    \rowcolor[HTML]{DDEBFF}
    2018  & Neural ODE \cite{NODE}  & ODE + RNN       & /      \\
    \rowcolor[HTML]{DDEBFF}
    2019  & ODE-RNN \cite{ode-rnn}  & ODE + RNN       & /      \\
    \rowcolor[HTML]{DDEBFF}
    2021  & Neural SDE \cite{nsde}  & ODE + GAN       & Stocks \\
    \rowcolor[HTML]{DDEBFF}
    2022  & GT-GAN \cite{gtgan}     & ODE + GAN       & Stocks \\
    \rowcolor[HTML]{DDEBFF}
    2023  & LS4 \cite{ls4}          & ODE + VAE       & /      \\
    \rowcolor[HTML]{DDEBFF}
    2024  & TimeLDM \cite{timeldm}  & Diffusion + VAE & Stocks \\
    \bottomrule
\end{tabular}}
\end{table}

\paragraph{Trading Performance}
Statistical accuracy does not guarantee financial profitability. 
We therefore simulate trading execution to evaluate economic utility.
Given the hourly profit-and-loss $\Delta V_t$ and simple return of equity $\Delta V_t/V_{t-1}$ at time $t$:
\begin{itemize}[nolistsep,left=2pt] %
  \item \keypoint{E5: Compound Annual Growth Rate (CAGR)} captures the annualized return based on equity growth, where $V_{0}$ and $V_{s}$ are the initial and final equity. It is calculated as:
  \begin{displaymath}
    \text{CAGR} = \big(\tfrac{V_{s}}{V_{0}}\big)^{8760/s} - 1.
  \end{displaymath}

  \item \keypoint{E6: Sharpe Ratio (SR)} is defined as:
  \begin{displaymath}
    \text{SR} = \tfrac{\mathbb{E}[\Delta V_t/V_{t-1}]}{\text{Std}(\Delta V_t/V_{t-1})} \cdot \sqrt{8760}.
  \end{displaymath}

\end{itemize}
These metrics quantify not only returns but also the risk profile of synthetic-data-driven trading strategies.

\paragraph{Risk Assessment Metrics}
Crypto markets are known for fat-tailed risks and sharp price swings. Generators that fail to reproduce these tail events can yield dangerously optimistic simulations.
Given profit-and-loss series $\Delta V_t$ and simple return of equity $\Delta V_t/V_{t-1}$:
\begin{itemize}[nolistsep,left=2pt] %
  \item \keypoint{E7: Maximum Drawdown (MDD)} is defined as:
  \begin{displaymath}
    \text{MDD} = \textstyle \max_{u \leq t} \big(\tfrac{V_u-V_t}{V_u}\big).
  \end{displaymath}
  
  \item \keypoint{E8: Value at Risk (VaR)} at 95\% confidence is defined as:
  \begin{displaymath}
    \text{VaR}_{0.95} = -\mathrm{Percentile}_{5\%}(\Delta V_t/V_{t-1}).
  \end{displaymath}

  \item \keypoint{E9: Expected Shortfall (ES)} at 95\% confidence is defined as:
  \begin{displaymath}
    \text{ES}_{0.95} = -\mathbb{E} \big[ (\Delta V_t/V_{t-1}) \mid (\Delta V_t/V_{t-1}) \leq -\text{VaR}_{0.95} \big].
  \end{displaymath}
\end{itemize}
VaR captures potential worst-day losses, while ES reveals mean loss beyond that threshold, offering a fuller picture of tail risk.

\paragraph{Efficiency}
Real-world crypto trading requires fast adaptation. Models must retrain frequently and generate data rapidly enough to integrate into live trading pipelines.
\begin{itemize}[nolistsep,left=2pt] %
  \item \keypoint{E10: Training Time} is the wall-clock time at which a TSG model is trained.

  \item \keypoint{E11: Inference Time} is the mean wall-clock time to generate or reconstruct one batch of data ($n$ assets $\times$ $s$ time steps).
\end{itemize}

\subsection{TSG Model Zoo}
\label{sec:benchmark:methods}

Generative models for time series aim to capture complex temporal dependencies and statistical patterns in sequential data.
As noted in \cite{Ang2023TSGBench,tsgm}, these models are typically categorized by their backbone architectures, such as VAEs, GANs, diffusion models, flow-based models, and mixed-type models, as summarized in Table \ref{tab:methods}.

Yet, nearly half of prior TSG studies have not evaluated their models in financial contexts.
Even among those that do, most focus narrowly on traditional markets, particularly equities (e.g., Google stock data in \cite{timegan}), offering limited insights for cryptocurrency applications. 
To bridge this gap, \textsf{CTBench} includes eight representative TSG models spanning all five methodological categories, selected to cover diverse architectures and modeling paradigms prevalent in recent literature~\cite{Ang2023TSGBench,tsgm}.

\paragraph{GAN-based Methods}
These methods \cite{cosci-gan, quant-gan,rtsgan,aec-gan} leverage adversarial training dynamics to generate realistic series.\footnote{GAN-based methods are used only in the cryptocurrency forecasting task, as GANs do not natively support reconstruction \cite{goodfellow2020generative, dumoulin2017adversarially}.}
They incorporate recurrent neural architectures and specialized attention mechanisms tailored to temporal dependencies.
\begin{itemize}[nolistsep,left=2pt] %
  \item \keypoint{M1: Quant-GAN \cite{quant-gan}} approximates a trading utility function, optimizing the generator for downstream profitability.
  \item \keypoint{M2: COSCI-GAN \cite{cosci-gan}} integrates causal self-attention and statistical conditioning to consider temporal order and cross-asset correlations.
\end{itemize}

\paragraph{VAE-based Methods}
These Methods use variational inference to capture both local and global temporal patterns \cite{timevae, timevqvae, crvae}. 
They have shown strong performance in general TSG tasks \cite{Ang2023TSGBench, bao2024towards}.
\begin{itemize}[nolistsep,left=2pt] %
  \item \keypoint{M3: TimeVAE \cite{timevae}} is a sequence-aware VAE with temporal convolutions, designed to capture both local and long-range dependencies in multivariate time series.

  \item \keypoint{M4: KoVAE \cite{kovae}} enhances TimeVAE by incorporating Koopman operator-based latent dynamics for smoother and more interpretable generation.
\end{itemize}

\paragraph{Diffusion-based Methods}
Diffusion models \cite{yuan2024diffusion, galib2024fide, chen2024sdformer, Li_Meng_Bi_Urnes_Chen_2025, naiman2024utilizing} progressively convert noise into structured data via iterative denoising, proving highly effective in modeling complex market dynamics.
\begin{itemize}[nolistsep,left=2pt] %
  \item \keypoint{M5: Diffusion-TS \cite{yuan2024diffusion}} is a score-based diffusion model that iteratively refines Gaussian noise into realistic trajectories, achieving state-of-the-art sample fidelity on financial data.
  
  \item \keypoint{M6: FIDE \cite{galib2024fide}} introduces factorized conditional diffusion with attention-driven score networks, enabling conditional generation based on market regimes or liquidity factors.
\end{itemize}

\paragraph{Flow-based Methods}
Flow-based methods \cite{fourier, hu2024fm} employ invertible transformations to model data distributions, ensuring exact likelihood estimation and efficient sampling.
\begin{itemize}[nolistsep,left=2pt] %
  \item \keypoint{M7: Fourier-Flow \cite{fourier}} uses frequency-domain coupling layers for invertible transformations, allowing fast sampling and exact likelihood computation while preserving periodic structures. 
\end{itemize}

\paragraph{Mixed-based Methods}
Hybrid models \cite{ls4, ode-rnn, gtgan} typically combine multiple modeling paradigms (e.g., ODEs and VAEs) to capture nuanced temporal dynamics and stochastic characteristics.
\begin{itemize}[nolistsep,left=2pt] %
  \item \keypoint{M8: LS4 \cite{ls4}} fuses deep state-space modeling with stochastic latent variables via variational inference, offering flexible and interpretable modeling of complex crypto market dynamics.
\end{itemize}

\section{Experiments}
\label{sec:expt}

\subsection{Experimental Setup}
\label{sec:expt:setup}

\paragraph{Datasets}
We employ the datasets \cite{binance_data} described in \sec{\ref{sec:benchmark:data}} for the experiments.
To simulate real-world deployment, we adopt a walk-forward rolling-window validation scheme, using 500 days of hourly data for training, and 30 or 15 days for testing on the Predictive Utility and Statistical Arbitrage tasks, respectively.
After each cycle, the window advances by the test period length, with models retrained. This process spans from January 2020 to December 2024, covering diverse market regimes.

\paragraph{Benchmark Configurations}
To isolate core TSG model performance, we assume zero trading fees by default in both Predictive Utility and Statistical Arbitrage tasks, enabling fair comparison of signal quality without interference from platform-specific costs. 
For the Statistical Arbitrage task, we also apply a 0.03\% trading fee, reflecting the fee level that a typical liquidity provider can achieve on major centralized exchanges \cite{zhang2023towards, winkel2023pricing, binance2025fees}, providing a more grounded evaluation of net profitability.

\paragraph{Trading Strategies}
For the Predictive Utility task, we employ three representative trading strategies in \sec{\ref{sec:benchmark:strategy}} to evaluate synthetic data across varied portfolio constructions.
In contrast, the Statistical Arbitrage task employs the mean-reversion strategy to isolate the model's ability to preserve exploitable residual structures.

\paragraph{TSG Methods}
We evaluate eight representative TSG models across five major families in \sec{\ref{sec:benchmark:methods}}. Hyperparameter settings follow published recommendations or are tuned for stable training.
\begin{itemize}[nolistsep,left=2pt] %
  \item \textbf{GAN-based:} 
  Quant-GAN adopts latent\_dim = 8, hidden\_dim = 80, gradient penalty $\lambda_{\text{gp}} = 10.0$, and critic steps $n_{\text{critic}} = 5$; 
  COSCI-GAN uses latent\_dim = 32, $\gamma = 5$, and $n_{\text{groups}} = 4$ with MLP-based central discriminators, as per \cite{cosci-gan}.
  
  \item \textbf{VAE-based:} 
  TimeVAE uses latent\_dim = 8 with stacked hidden layers of 50, 100, and 200 units; 
  KoVAE follows \cite{kovae}, setting $W_{\mathrm{KL}} = 0.009$ and $W_{\mathrm{PRED}} = 0.03$ for KL and auxiliary loss terms.
  
  \item \textbf{Diffusion-based:}
  Diffusion-TS uses 1000 timesteps, 3 encoder layers, 6 decoder layers, and $d_{\text{model}} = 64$;  
  FIDE applies 1000 steps, hidden\_dim = 64, 8 layers, and $\sigma = 0.05$.

  \item \textbf{Flow-based:}
  Fourier-Flow incorporates DFT-based coupling layers with hidden\_size = 128 and 3 flow layers.

  \item \textbf{Mixed-type:} 
  LS4 employs hidden\_dim = 6, latent\_dim = 8, and a batch size of 512.
\end{itemize}

\paragraph{Evaluation Measures}
We adopt the twelve metrics detailed in \sec{\ref{sec:benchmark:eval}}, thereby scoring each model on forecasting accuracy, rank correlation, trading profitability, tail risk, and computational efficiency.

\paragraph{Experiments Environments}
All experiments are conducted on a machine equipped with an Intel\textsuperscript{\textregistered} Xeon\textsuperscript{\textregistered} Platinum 8480C @3.80GHz, 64 GB RAM, and an NVIDIA H100 GPU.

\begin{figure*}[t]
\centering
\captionsetup{skip=0.75em,belowskip=0.0em}%
\includegraphics[width=0.99\textwidth]{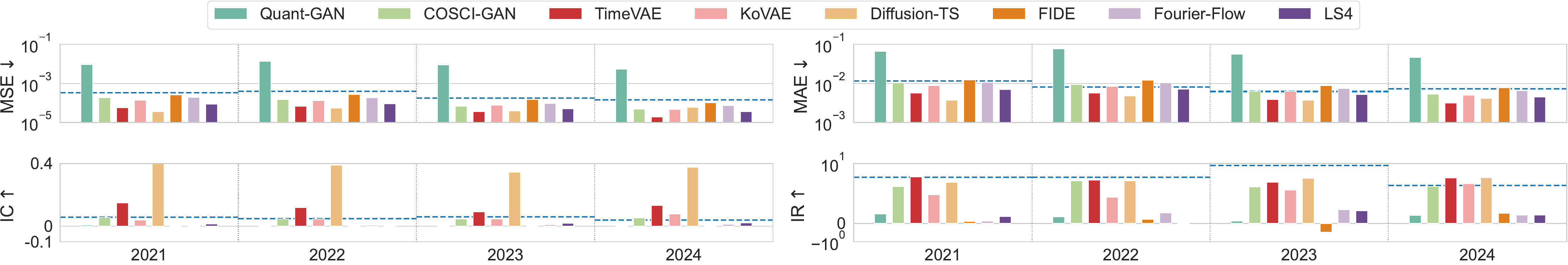}
\vspace{-0.25em}%
\caption{Annual forecasting performance of TSG methods on the Predictive Utility task.}%
\label{fig:gfb_forecasting}%
\vspace{-0.5em}
\end{figure*}

\begin{figure*}[t]
\centering
\captionsetup{skip=0.75em,belowskip=0.0em}%
\includegraphics[width=0.99\textwidth]{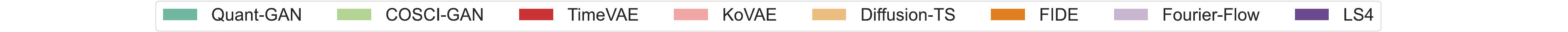} \\
\includegraphics[width=0.99\textwidth]{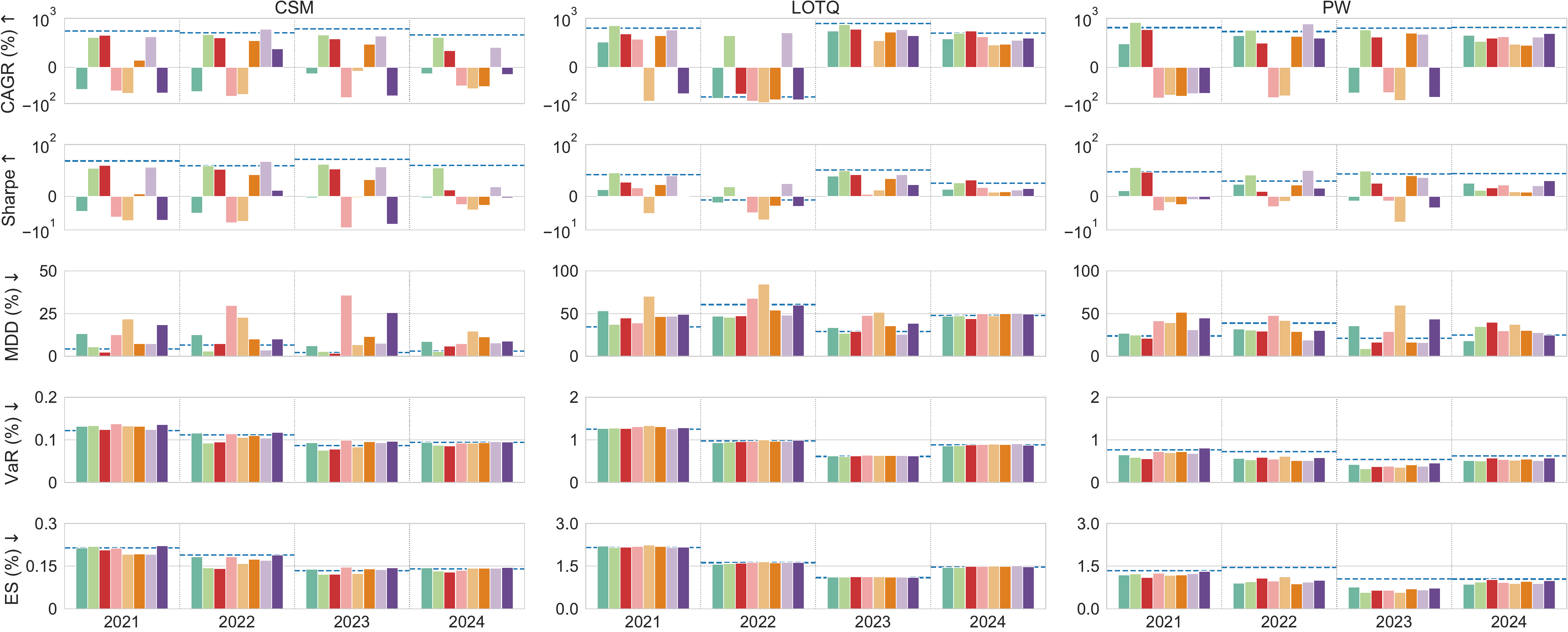}
\vspace{-0.25em}%
\caption{Annual trading performance of TSG methods on the Predictive Utility task.}%
\label{fig:gfb_trading}%
\vspace{-0.5em}
\end{figure*}

\subsection{Predictive Utility Task}
\label{sec:expt:generation_experiments}

Figures~\ref{fig:gfb_forecasting} and~\ref{fig:gfb_trading} show the year-wise performance of TSG models from 2021 to 2024, highlighting forecasting accuracy and trading effectiveness, respectively.
The blue dashed line denotes the baseline using real data (without TSG), whose strong performance underscores the effectiveness of our feature extraction pipeline (\sec{\ref{sec:benchmark:data}}).

\paragraph{Annual Predictive Utility Analysis}
In the 2021 bull market, Diffusion-TS leads in predictive accuracy, suggesting that its score-based denoising mechanism effectively captures transient momentum. 
However, this statistical strength does not yield profitable trading--its negative CAGR and low Sharpe ratio highlight an accuracy–alpha gap, where fidelity suppresses the volatility essential for directional gains.
In contrast, TimeVAE strikes a compelling balance, delivering solid forecasting accuracy and robust returns, likely due to its variational bottleneck, which filters noise while preserving exploitable variance.
COSCI-GAN thrives under trend-sensitive strategies such as LOTQ and PW, producing promising CAGRs and Sharpe ratios above five. While its IC and IR scores are modest, the model clearly amplifies alpha in bullish conditions. 
Flow-based models, notably Fourier-Flow, exhibit a conservative profile with moderate rank fidelity, stable but subdued returns, and minimal drawdowns, which indicates that invertibility might introduce useful constraints on overfitting.

In the volatile 2022 market, all models see moderate declines in forecasting accuracy. 
Yet, TimeVAE remains robust, achieving positive Sharpe ratios across strategies.
Diffusion-TS, despite leading in error-based metrics, suffers from prolonged drawdowns, underscoring its vulnerability to directional reversals. 
COSCI-GAN yields high CAGR under CSM with shallow drawdowns, suggesting effective exploitation of volatility-induced dispersion.
LS4 prioritizes risk control over ranking precision, serving as a practical hedge in chaotic regimes.

In 2023's consolidation phase, prediction errors narrow, but trading outcomes diverge sharply. 
Trend-reliant models like COSCI-GAN falter, while dispersion-sensitive models such as TimeVAE and Fourier-Flow maintain high Sharpe ratios. 
Notably, Fourier-Flow excels with low tail risk and strong risk-adjusted returns, showcasing its strength in frequency-preserving synthesis under range-bound conditions.

By 2024, in a mean-reverting regime, both predictive accuracy and profitability contract further.
This low-signal setting challenges model generalization. TimeVAE maintains marginal profitability, but most models fail to generate consistent returns, highlighting the limits of fidelity-focused generation in environments with sparse, fleeting alpha opportunities.

\begin{figure*}[t]
\centering
\captionsetup{skip=0.75em,belowskip=0.0em}%
\includegraphics[width=0.80\textwidth]{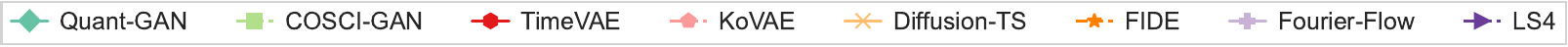}
\\
\subfigure[2021.]{%
  \label{fig:gfb_ranking:2021}%
\includegraphics[width=0.245\textwidth]{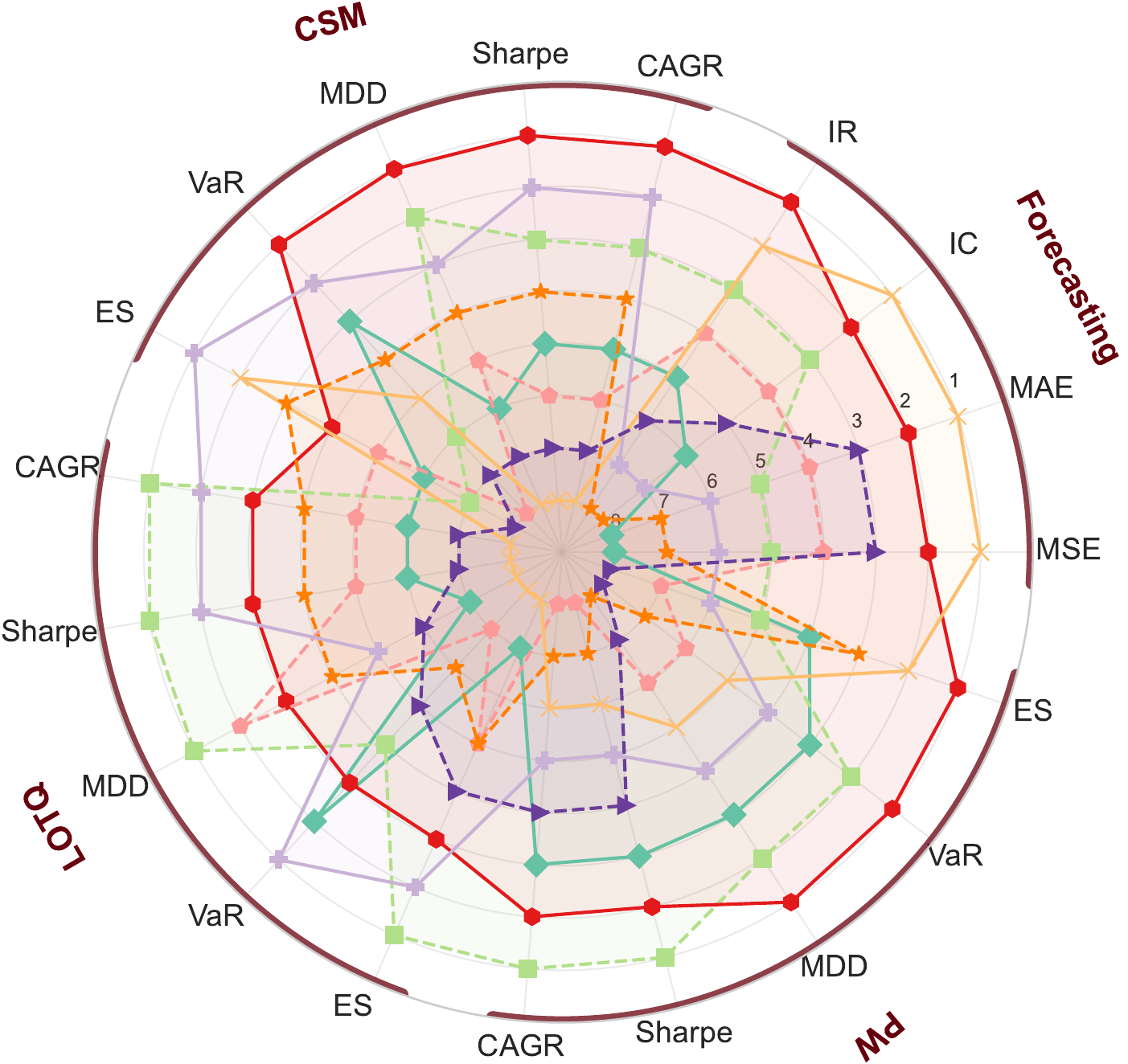}}
\subfigure[2022.]{%
  \label{fig:gfb_ranking:2022}%
\includegraphics[width=0.245\textwidth]{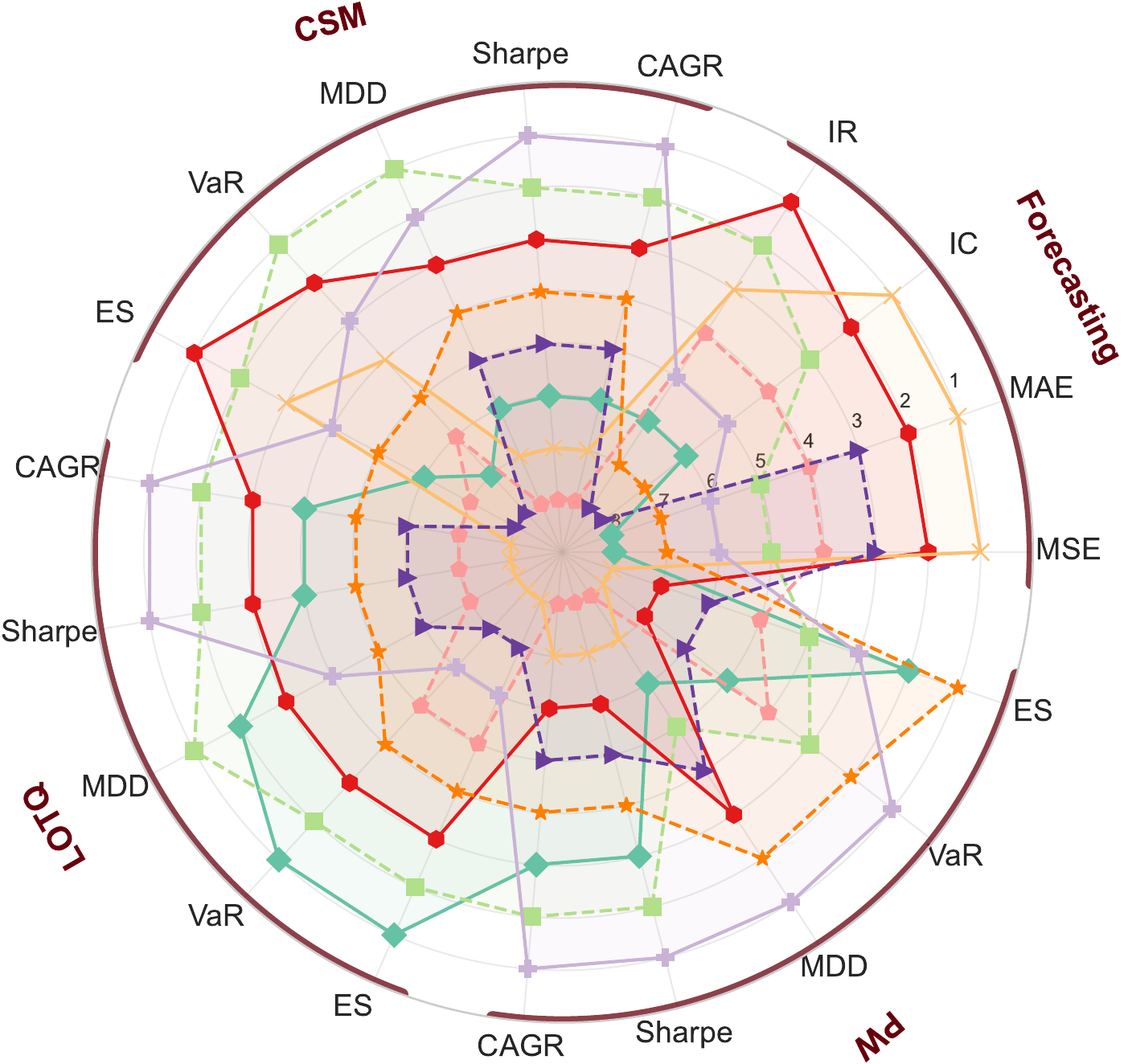}}
\subfigure[2023.]{%
  \label{fig:gfb_ranking:2023}%
\includegraphics[width=0.245\textwidth]{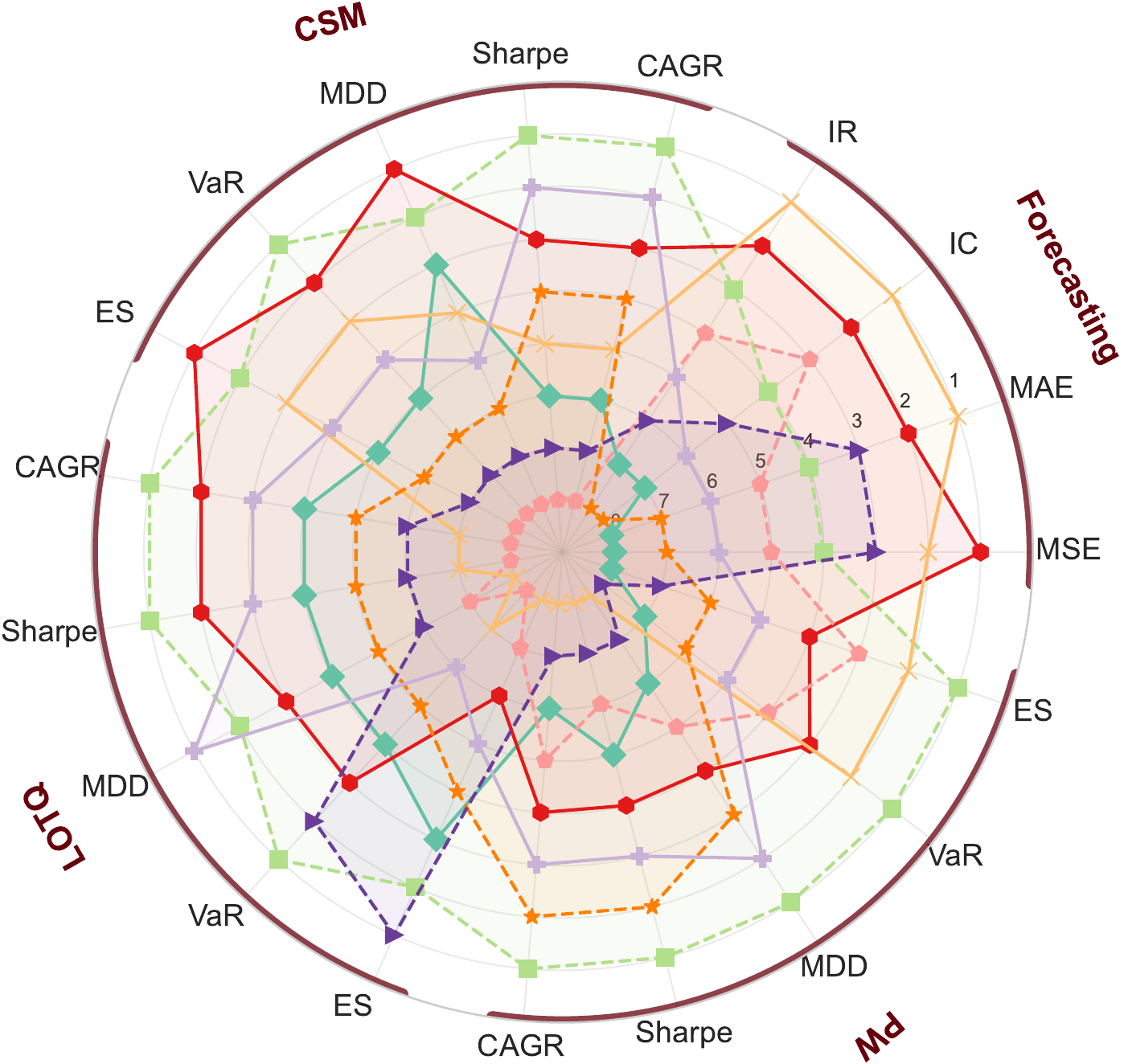}}
\subfigure[2024.]{%
  \label{fig:gfb_ranking:2024}%
\includegraphics[width=0.245\textwidth]{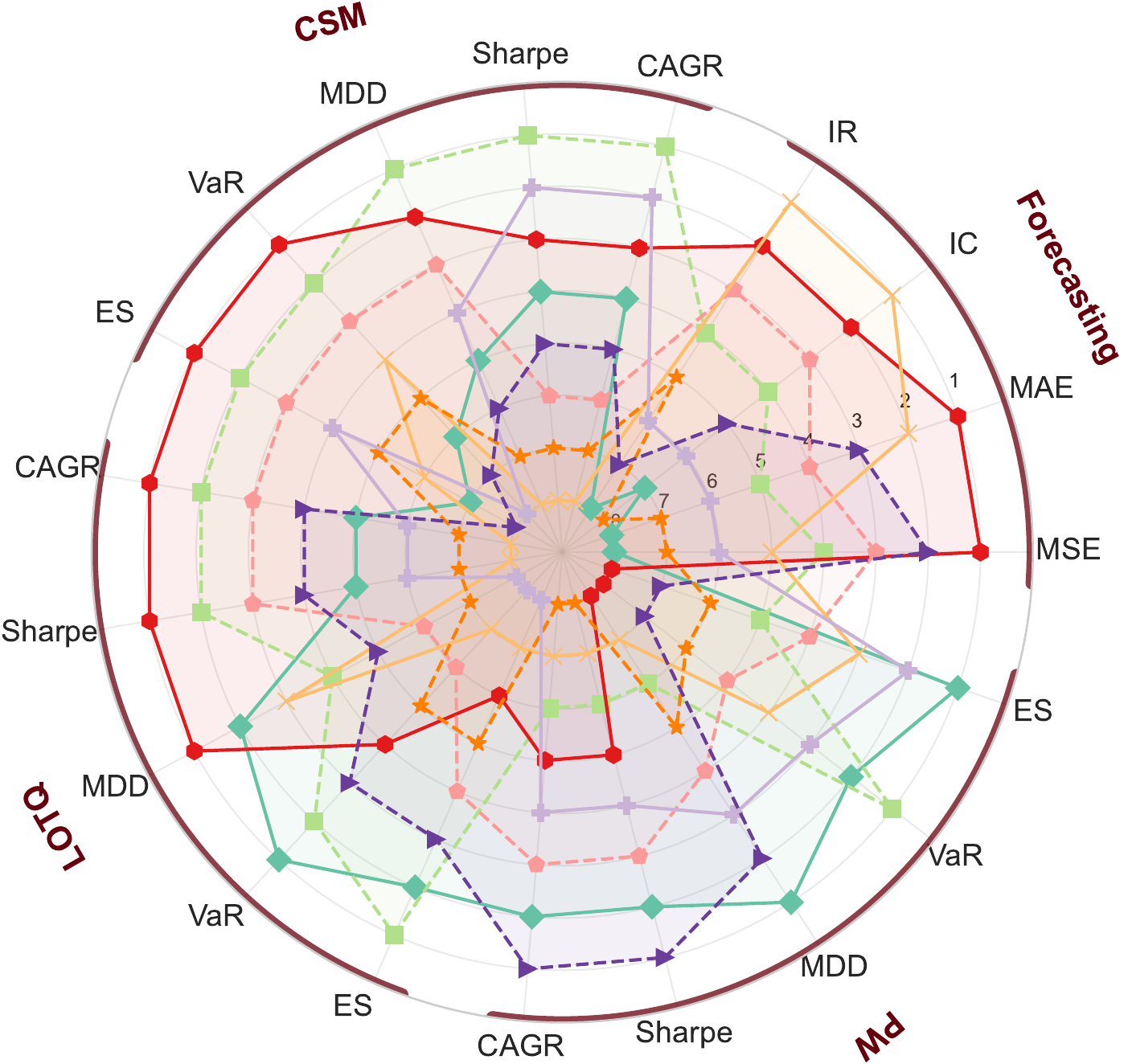}}
\vspace{-0.5em}%
\caption{Rankings of TSG models on the Predictive Utility task.}%
\label{fig:gfb_ranking}%
\vspace{-0.5em}
\end{figure*}

\begin{figure*}[t]
  \centering
  \includegraphics[width=0.99\textwidth]{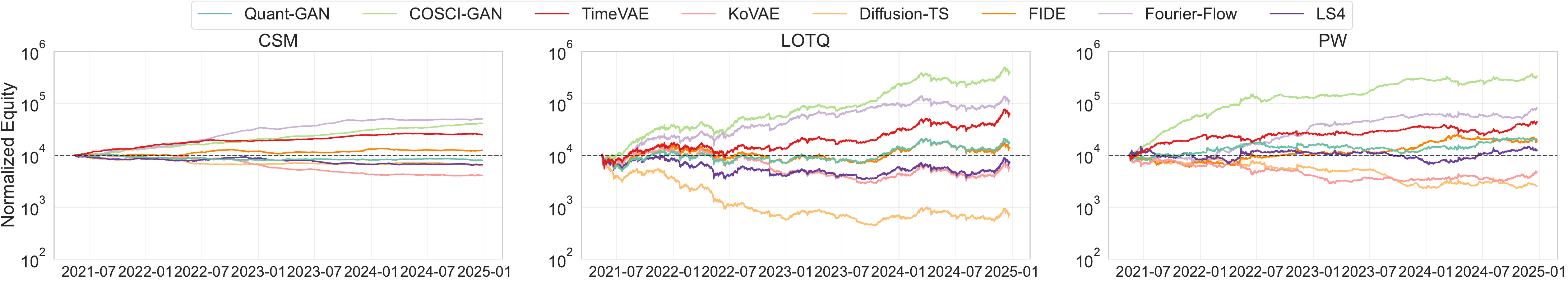}
  \vspace{-0.5em}
  \caption{Simulated growth curves of a \$10,000 investment over four years under three trading strategies.}
  \label{fig:exp-generate_equity_curves}
  \vspace{-0.5em}
\end{figure*}

\paragraph{Ranking Analysis}
Figure~\ref{fig:gfb_ranking} summarizes model performance via radar plots, revealing three key patterns:
(1) Diffusion-TS consistently ranks highest in forecasting metrics but lags in trading performance, highlighting a classic case of economic inefficiency in high-fidelity generation.
(2) TimeVAE and COSCI-GAN exhibit regime-dependent strengths: TimeVAE excels in stable or mean-reverting markets, likely due to its regularization, while COSCI-GAN thrives in volatile, directional regimes where high variance amplifies trend signals.
(3) Fourier-Flow maintains stable mid-to-high rankings across all metrics, emerging as a robust all-weather model suitable for risk-managed deployment.

Together, these findings underscore a core insight: \emph{low reconstruction or prediction error does not guarantee trading success.} 
Over-regularized models like Diffusion-TS or LS4 may suppress alpha-rich variance, diminishing profitability. In contrast, models that retain structural noise or tail behavior, such as TimeVAE and COSCI-GAN, offer greater real-world utility. 
Therefore, Effective model selection requires regime awareness and alignment with strategy goals. 
Prioritizing synthetic fidelity alone is insufficient; deploying \textsf{CTBench} successfully demands a balanced view of both predictive realism and financial viability.

\paragraph{Equity Curve Dynamics}
Figure~\ref{fig:exp-generate_equity_curves} shows log-scaled equity curves (starting from \$10,000) for each TSG model under three trading strategies from 2021 to 2024, illustrating cumulative returns and how model inductive biases interact with market regimes.

Under \textbf{CSM}, COSCI-GAN and TimeVAE achieve steady gains by preserving rank order and alpha, though they cap upside by dampening extreme winners.
In contrast, Diffusion-TS and FIDE steadily decline, as denoising suppresses volatility and undermines long–short execution.
Under \textbf{LOTQ}, COSCI-GAN emerges as the clear leader, likely benefiting from adversarially enhanced right-tail signals that capture strong directional gains. 
TimeVAE and Fourier-Flow maintain modest, stable growth, while Diffusion-TS continues to falter due to loss of rare but critical upward spikes.
Under \textbf{PW}, which rewards consistent pairwise ranking, COSCI-GAN again dominates.
TimeVAE and Fourier-Flow show smooth compounding, reflecting robust generalization from well-regularized latent spaces. LS4, by contrast, remains largely flat across all strategies, indicating its conservative design acts more like a low-beta portfolio.
These dynamics underscore the importance of aligning model characteristics with strategy needs, particularly in volatile markets.

\begin{figure*}[t]
  \centering
  \includegraphics[width=0.99\textwidth]{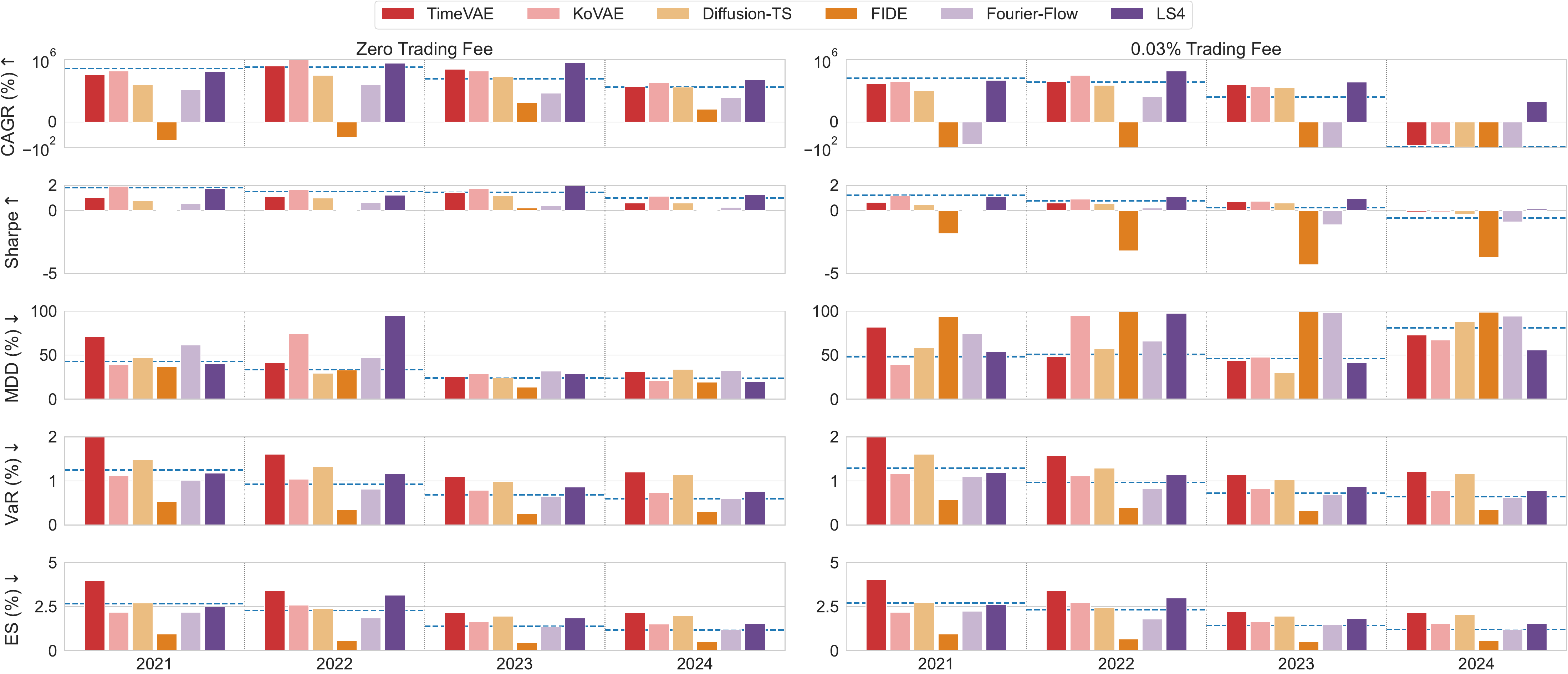}
  \vspace{-0.5em}
  \caption{Annual performance of TSG methods on the Statistical Arbitrage task.}%
  \label{fig:rtb}%
  \vspace{-0.25em}
\end{figure*}

\begin{figure*}[t]
\centering
\captionsetup{skip=0.75em,belowskip=0.0em}%
\includegraphics[width=0.80\textwidth]{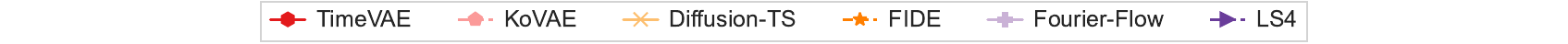}
\\
\subfigure[2021.]{%
  \label{fig:rtb_ranking:2021}%
\includegraphics[width=0.245\textwidth]{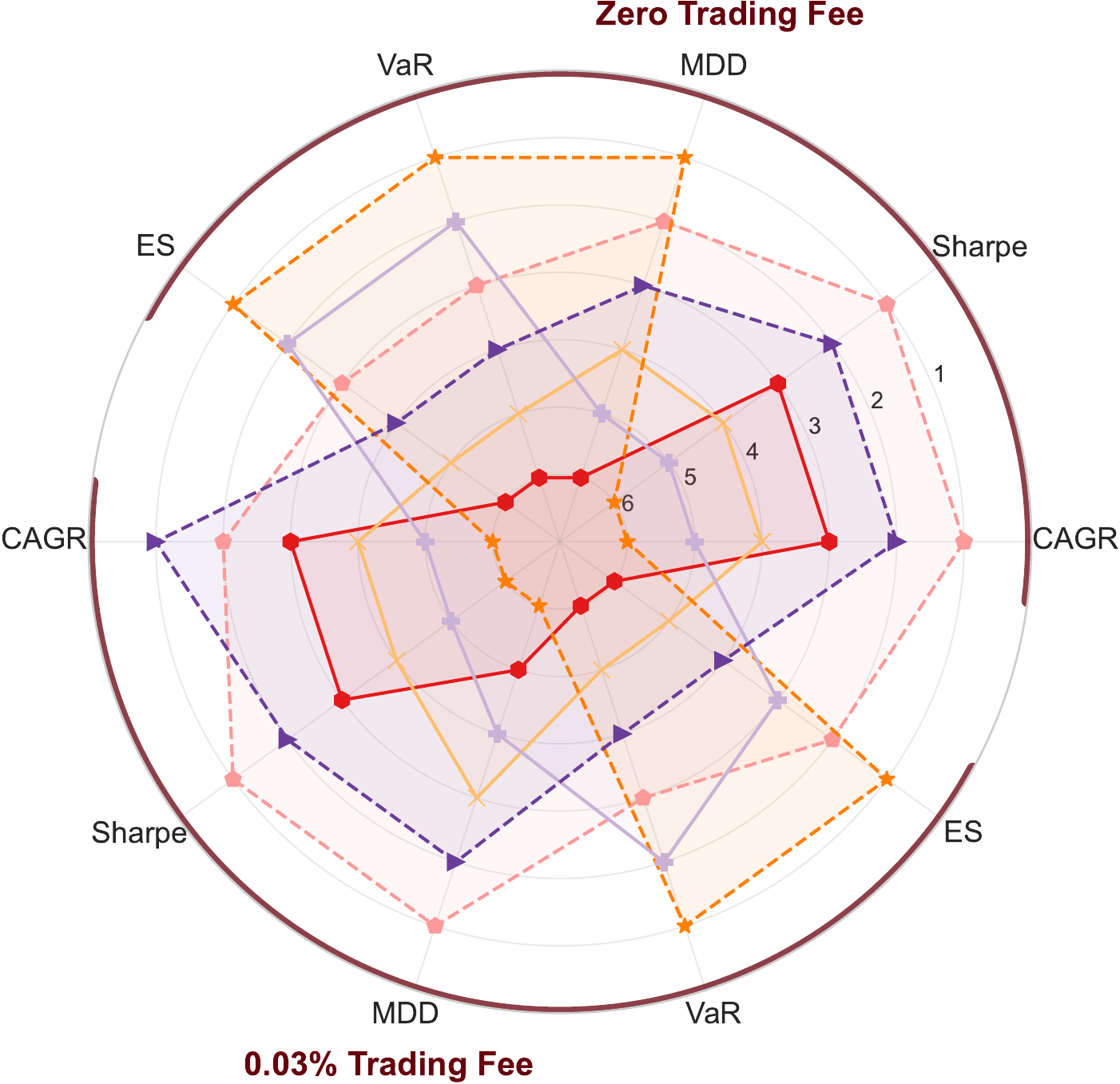}}
\subfigure[2022.]{%
  \label{fig:rtb_ranking:2022}%
\includegraphics[width=0.245\textwidth]{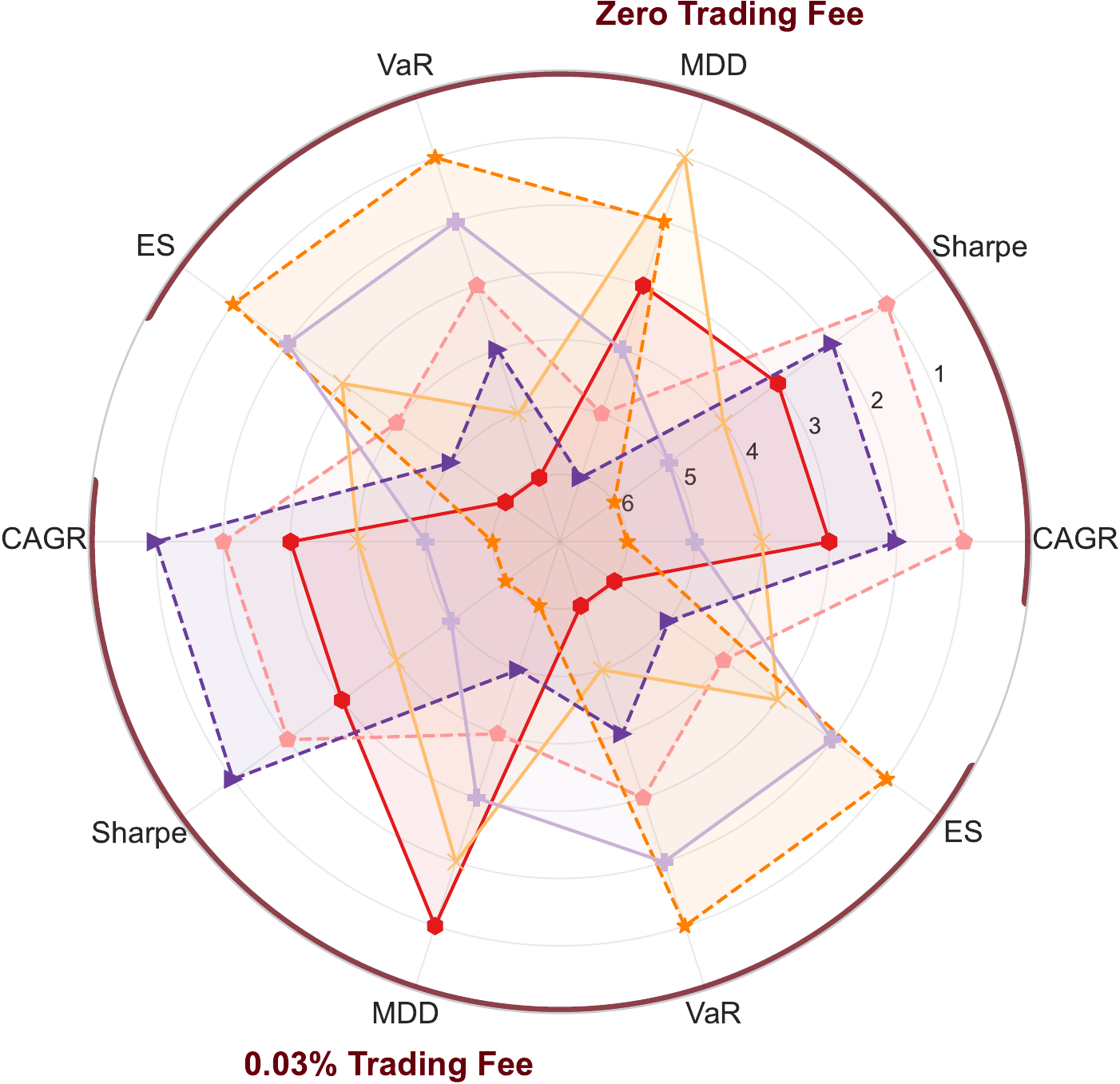}}
\subfigure[2023.]{%
  \label{fig:rtb_ranking:2023}%
\includegraphics[width=0.245\textwidth]{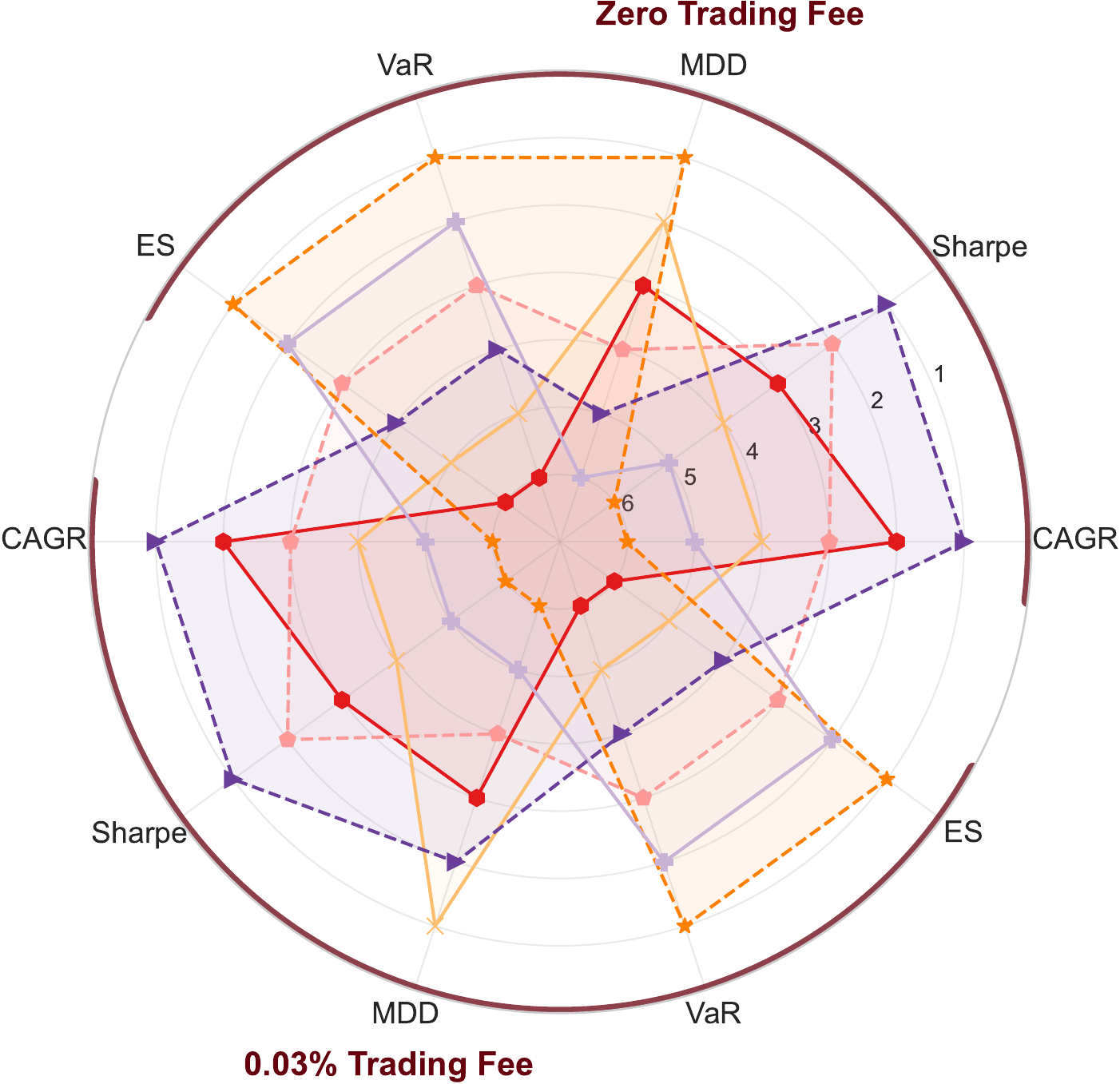}}
\subfigure[2024.]{%
  \label{fig:rtb_ranking:2024}%
\includegraphics[width=0.245\textwidth]{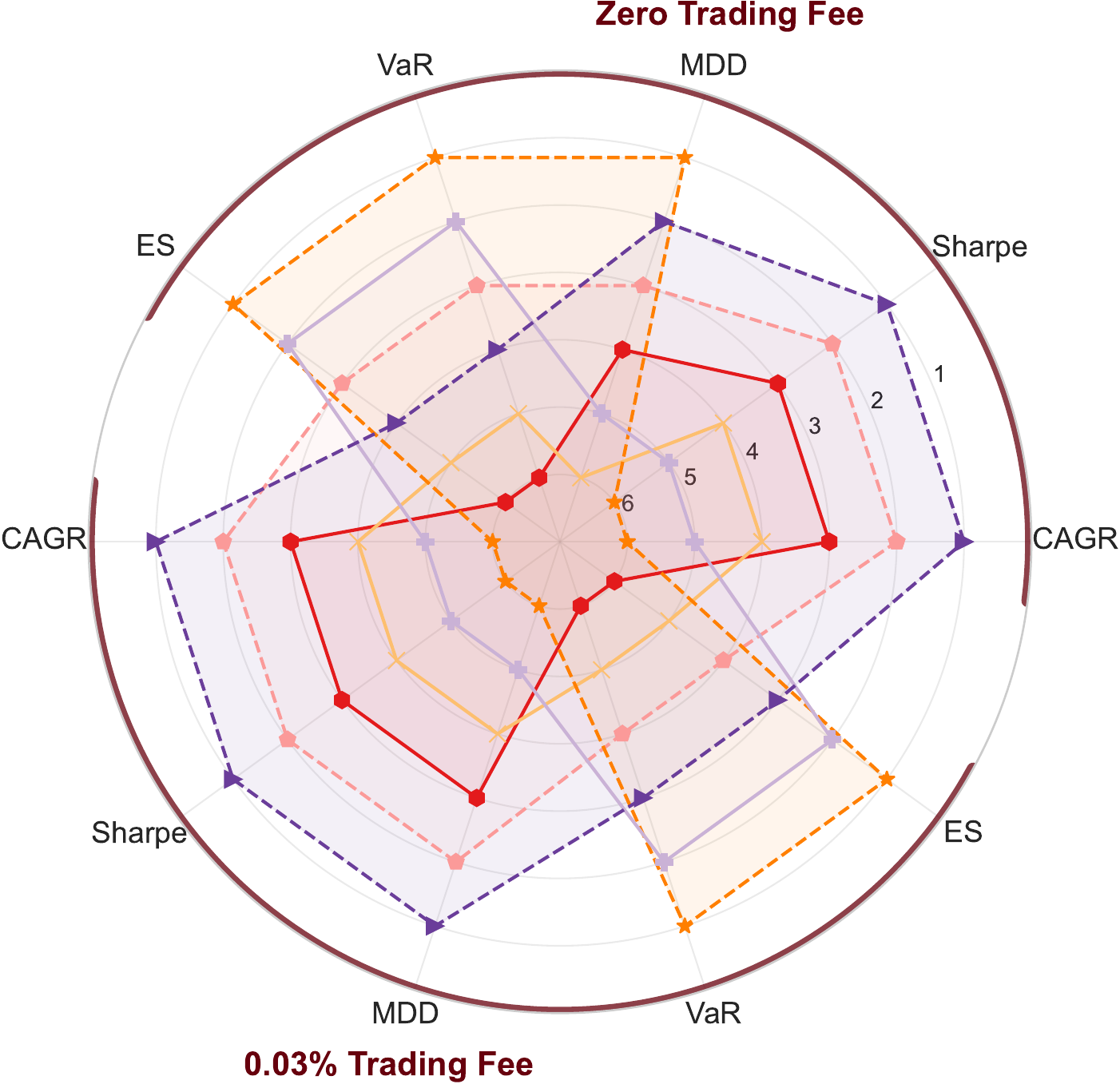}}
\vspace{-0.5em}%
\caption{Rankings of TSG models on the Statistical Arbitrage task.}%
\label{fig:rtb_ranking}%
\vspace{-0.5em}
\end{figure*}

\subsection{Statistical Arbitrage Task}
\label{sec:results:reconstruction_benchmark}

Figure~\ref{fig:rtb} reports the annualized trading performance and risk metrics of various TSG models, under both idealized and realistic trading scenarios.
The blue dashed line shows a baseline using a Principal Component Analysis (PCA) model calibrated on $\bm{R}_{\text{train}}$, reflecting a classical approach used by statistical arbitrage desks and serving as a reference point for evaluating TSG models.

\paragraph{Annual Performance Analysis}
Across the four years, while all models suffer a drop in profitability when trading fees are introduced, the extent of the degradation varies with each model's trading frequency and volume; those that trade most often incur the greatest drag, while smoother, lower‑turnover strategies retain more of their gains.
Among the TSG models, KoVAE and LS4 consistently rank near the top in terms of annual returns, albeit through very different risk postures. 
In the crisis-like environment of 2022, KoVAE records the highest CAGR but incurs substantial drawdowns and a moderate Sharpe ratio, indicating large but mean-reverting profit swings. 
In contrast, LS4 shines in 2023, delivering both the best CAGR and Sharpe ratio of the year while maintaining a relatively contained MDD. After accounting for trading fees, both models retain top-tier positions, but their raw CAGRs shrink, illustrating that even alpha-rich residuals require careful cost control to remain viable.
TimeVAE and Diffusion-TS form a second tier of models that trade off headline returns for improved risk-adjusted stability. While they seldom lead in CAGR, their Sharpe ratios remain positive and relatively fee-resistant. However, both of them occasionally exhibit large tail risk, as reflected in elevated VaR and ES levels, especially in 2021 and 2024, which drag down their overall risk-return efficiency.
FIDE, on the other hand, delivers near-zero or negative CAGRs and Sharpe ratios across all years, but it repeatedly achieves the lowest VaR/ES and often the smallest MDD. In other words, FIDE reconstructs residuals that might be ``too clean'' to trade.
Fourier-Flow also underperforms in returns while failing to consistently control drawdowns, indicating that exact-likelihood flow models might smooth out high-frequency noise but do not necessarily isolate tradable, mean-reverting components.

\paragraph{Ranking Analysis}
The radar plots in Figure~\ref{fig:rtb_ranking} illustrate distinct geometric patterns, revealing diverse model behaviors under varying market conditions.
Models such as KoVAE and LS4 display polygons that sharply bulge toward the CAGR and Sharpe Ratio axes, signaling strong returns but simultaneously cave in on the risk axes, particularly in turbulent periods. 
In contrast, FIDE produces the inverse shape: risk metrics are tightly controlled, but return metrics collapse, reaffirming its capital-preserving but alpha-deficient nature.
TimeVAE and Diffusion-TS exhibit more balanced polygonal profiles, with no dominant vertices but also no significant collapses. These shapes suggest regime-agnostic robustness, models that might not excel in any single dimension but offer resilience across diverse conditions.
One of the more subtle yet practically meaningful insights lies in the transformation of these rank profiles when fees are introduced. Although the overall topology of each polygon remains consistent, the rank distances compress. High-turnover models such as KoVAE drop multiple Sharpe positions under fee scenarios, while smoother models like TimeVAE and Diffusion-TS show smaller rank erosion. This implies that smoother residual signals might naturally induce lower turnover, yielding better fee-adjusted outcomes.
Moreover, year-over-year changes in polygon shape further expose model-specific regime sensitivities. For instance, LS4 exhibits dramatic expansion in CAGR during 2023 but contracts sharply on MDD in 2022. Conversely, KoVAE peaks during turbulent regimes but underperforms in calmer periods.

\begin{figure}[t]
  \centering
  \includegraphics[width=0.99\columnwidth]{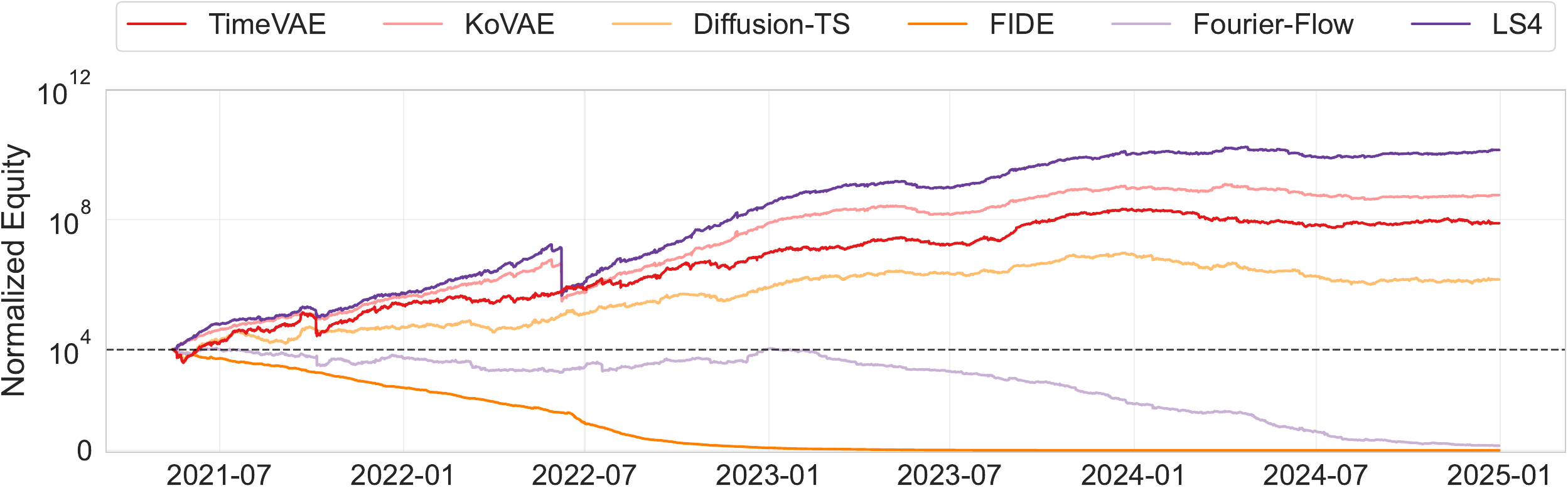}
  \vspace{-0.5em}
  \caption{Simulated growth curves of a \$10,000 investment for the Statistical Arbitrage task (with 0.03\% fee).}
  \label{fig:exp-recon_equity_curves}
  \vspace{-0.5em}
\end{figure}

\begin{table*}[htbp]
\centering
\small
\caption{Scenario-based recommendations for selecting TSG models in cryptocurrency markets.}
\label{tab:recommendation}%
\vspace{-1.0em}
\begin{tabular}{p{0.15\textwidth}p{0.23\textwidth}p{0.56\textwidth}}
\toprule
\rowcolor[HTML]{FFF2CC} \textbf{Scenario} & \textbf{Recommended TSG Models} & \textbf{Rationale} \\
\midrule
Trend-following / Directional Markets & COSCI-GAN, KoVAE & COSCI-GAN amplifies trend and dispersion; KoVAE offers alpha with higher drawdowns \\
\rowcolor[HTML]{DDEBFF} Mean-reverting / Range-bound Regimes & TimeVAE, Fourier-Flow, Diffusion-TS & TimeVAE/Fourier-Flow provide balance; Diffusion-TS preserves rank order \\
Fee-sensitive / Low-turnover Settings & TimeVAE, Diffusion-TS & Smooth residuals, stable Sharpe under transaction costs \\
\rowcolor[HTML]{DDEBFF} Risk Tolerance / Portfolio Design & KoVAE, LS4, TimeVAE, Diffusion-TS, FIDE & KoVAE/LS4 maximize returns with risk; TimeVAE/Diffusion-TS balance Sharpe and drawdown; FIDE is defensive \\
Deployment Efficiency & TimeVAE, LS4 & Fast retraining and low-latency inference; diffusion models better suited for offline use \\
\bottomrule
\end{tabular}
\end{table*}%

\paragraph{Equity Curve Dynamics}
Figure~\ref{fig:exp-recon_equity_curves} illustrates the equity curves under the Statistical Arbitrage task, initialized at \$10{,}000, with 0.03\% trading fees. 
At the top end, LS4 compounds almost monotonically, highlighting its superior fee resilience, and is punctuated by two staircase‑like surges in mid‑2022 and early 2023. 
This suggests that its latent‑switching mechanism excels at locking onto regime shifts rather than simply reacting to incremental mean‑reversion signals.
KoVAE follows with a similarly convex equity curve, initially smooth and robust with shallow drawdowns until late 2023, before growth tapers off in the more chaotic 2024 environment. TimeVAE shows steady gains through 2022, plateau in mid-2023, and drift sideways or slightly downward into 2024. This reflects its reliance on residual signals that are strong when cross-sectional dispersion is high but become increasingly exhausted as alpha opportunities compress. 
Diffusion-TS delivers the stable curve with minimal drawdowns, albeit with the lowest terminal return among viable models, consistent with its earlier characterization as a fee-resilient, risk-balanced generator. 
In contrast, FIDE collapses early, suggesting that its residuals are possibly over-regularized to the point of eliminating tradable structure. 
At the same time, Fourier-Flow bleeds capital slowly but persistently after mid-2022, likely due to over-smoothed residual patterns that incur persistent negative carry. 
Taken together, these dynamics emphasize that TSG models should balance fidelity and dispersion with regime adaptability to produce robust and economically viable equity trajectories.

\begin{figure}[t]
  \centering
  \includegraphics[width=0.99\columnwidth]{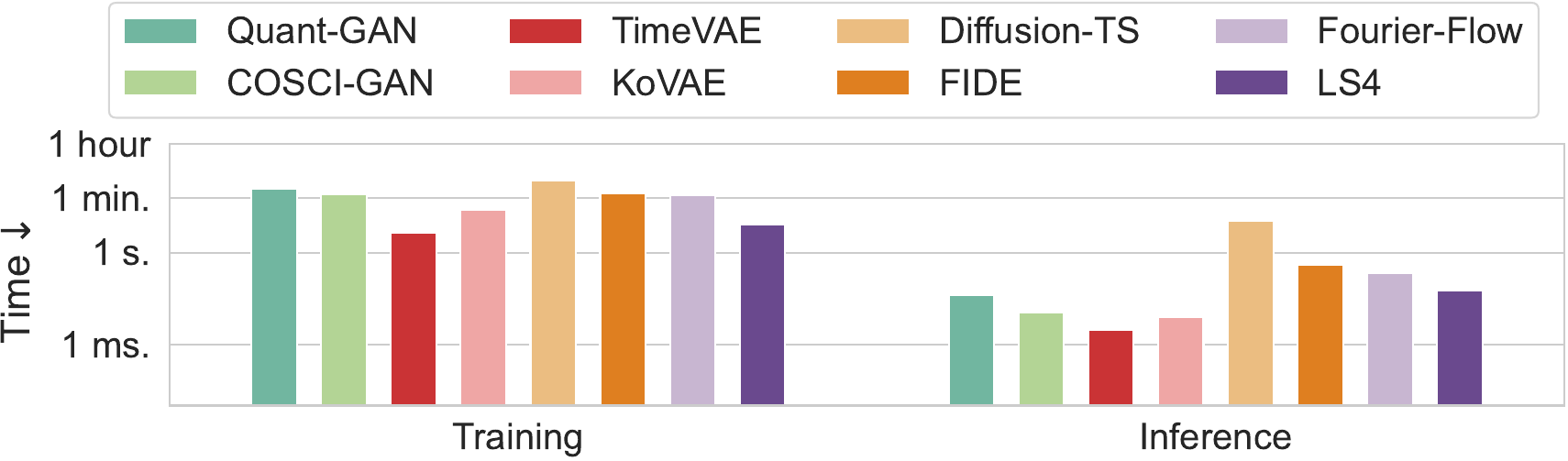}
  \vspace{-0.5em}%
  \caption{Training and inference time of TSG methods.}
  \label{fig:efficiency}%
  \vspace{-0.5em}
\end{figure}

\subsection{Efficiency}
\label{sec:results:efficiency}

Lastly, we compare the training and inference times of all TSG models in Figure~\ref{fig:efficiency}. 
VAE-based models stand out as the most computationally efficient. In particular, TimeVAE completes training in under a minute and achieves sub-second inference latency. This makes it especially attractive for real-time applications such as online data augmentation, low-latency strategy backtesting, and high-frequency retraining in rapidly evolving markets. 
GAN-based models offer moderate efficiency; while COSCI-GAN maintains a balanced runtime cost across both phases, Quant-GAN suffers from relatively high training time without commensurate improvements in generation speed. 
Diffusion-based models are the most computationally intensive, with Diffusion-TS incurring the longest training and inference durations due to its iterative denoising pipeline, and FIDE offering only marginal improvements. 
As such, despite their superior performance on fidelity and risk-return, they might be more suitable for offline use cases or environments with abundant compute resources.
Flow-based and mixed-type models sit between VAE and diffusion models. This makes them viable when likelihood calibration is essential, but latency is not a primary concern.

\subsection{Recommendations}
\label{sec:results:recommendations}

Our findings reveal a four-way trade-off among TSG model families:
(1) VAE-based models ensure stable reconstruction but might under-react to fast-changing regimes. 
(2) GAN-based approaches extract trend alpha but suffer from volatility-induced instability.
(3) Diffusion models handle regime clustering and fat tails well, but degrade under low signal regimes.
(4) Flow-based models prioritize likelihood but offer limited utility, while mixed-type ones are efficient but inconsistent in risk–return.

Based on these findings, Table \ref{tab:recommendation} distills them into actionable recommendations for the end-users.
These recommendations enable practitioners to align model selection with specific market conditions, strategic intents, and operational constraints. 
Importantly, the optimal use of TSG models in crypto is not a ``one-model-fits-all'' solution. Instead, users should: 
(1) diagnose their market regime, alpha source, and operational constraints, 
(2) select a TSG model whose inductive bias amplifies the desired structure without destroying tradability, and 
(3) evaluate it with a task–metric combination that mirrors the production objective. 
\textsf{CTBench}'s dual-task design and evaluation suite provide precisely this decision surface.

\section{Conclusion and Future Work}
\label{sec:conclusions}

In this paper, we introduce \textsf{CTBench}, the first benchmark tailored for TSG in cryptocurrency markets. 
\textsf{CTBench} integrates a curated high-frequency crypto dataset, a dual-task evaluation framework encompassing Predictive Utility and Statistical Arbitrage, and a rich suite of financial metrics designed to assess both statistical fidelity and real-world viability. 
Through extensive empirical analysis, we uncover critical trade-offs across TSG families and offer practical guidance for deploying models under diverse market conditions.

As a collaborative resource,  \textsf{CTBench} aims to foster rigorous evaluation and drive innovation in crypto time series modeling. 
Moving forward, we plan to expand \textsf{CTBench} by incorporating new tokens, extending to cross-exchange data, and integrating more advanced TSG architectures. 
We are also exploring model ensembling and regime-aware switching to improve robustness and performance consistency. 
To further streamline experimentation, we intend to support automated evaluation and hyperparameter tuning, enhancing both efficiency and usability.


\balance
\bibliographystyle{ACM-Reference-Format}
\bibliography{main}

\end{document}